\documentclass[10pt,journal,final]{IEEEtran}

\ifCLASSINFOpdf

\else

\fi

\usepackage{mathrsfs}
\usepackage{amsmath}
\usepackage{amsfonts}
\usepackage{amsthm}
\usepackage{amssymb}
\usepackage{graphicx}
\usepackage{subfigure}
\usepackage{indentfirst}
\usepackage{array}
\usepackage{epstopdf}
\usepackage{cite}
\usepackage{mathrsfs}
\usepackage{enumerate}
\usepackage{bm}
\usepackage{dsfont}
\usepackage[linesnumbered,ruled,vlined]{algorithm2e}
\usepackage{pifont}
\usepackage{setspace}
\usepackage{multirow}
\usepackage{verbatim}
\usepackage{color}
\usepackage{stfloats}
\usepackage{makecell}
\usepackage{cancel}
\usepackage{multicol}
\usepackage{adjustbox}

\usepackage[dvipsnames]{xcolor}
\usepackage[linkcolor=blue,citecolor=blue,colorlinks=true,urlcolor=blue]{hyperref}
\SetKwComment{Comment}{//}{}
\SetKwBlock{Begin}{Function}{end}
\SetKwInput{KwInput}{Input}                
\SetKwInput{KwOutput}{Output}              

\def\PPhi{{\boldsymbol \Phi}}
\def\PPsi{{\boldsymbol \Psi}}

\newcommand{\x}{{\boldsymbol x}}
\newcommand{\y}{{\boldsymbol y}}
\newcommand{\z}{{\boldsymbol z}}
\newcommand{\e}{{\boldsymbol e}}
\newcommand{\E}{{\boldsymbol E}}

\begin{document}

\title{SoundSpring: Loss-Resilient Audio Transceiver with Dual-Functional Masked Language Modeling}

\author{Shengshi Yao,~\IEEEmembership{Member, IEEE},
        Jincheng Dai,~\IEEEmembership{Member, IEEE},
        Xiaoqi Qin,~\IEEEmembership{Senior Member, IEEE}, \\
        Sixian Wang,~\IEEEmembership{Member,~IEEE},
        Siye Wang,~\IEEEmembership{Member, IEEE},
        Kai Niu,~\IEEEmembership{Member, IEEE},
        and Ping Zhang,~\IEEEmembership{Fellow, IEEE}

\thanks{This work was supported in part by the Beijing Natural Science Foundation under Grant L232047 and Grant 4222012, in part by the National Natural Science Foundation of China under Grant 62321001, Grant 62293481, Grant 62371063, and Grant 92267301, in part by Program for Youth Innovative Research Team of BUPT No. 2023YQTD02. \emph{(Corresponding author: Jincheng Dai)}}

\thanks{Shengshi Yao, Jincheng Dai, Sixian Wang, Siye Wang, and Kai Niu are with the Key Laboratory of Universal Wireless Communications, Ministry of Education, Beijing University of Posts and Telecommunications, Beijing 100876, China. (Corresponding author email: daijincheng@bupt.edu.cn)}

\thanks{Xiaoqi Qin, and Ping Zhang are with the State Key Laboratory of Networking and Switching Technology, Beijing University of Posts and Telecommunications, Beijing 100876, China.}

\thanks{Our code is available at~\url{https://github.com/semcomm/ResilientAudio}.}

}

\maketitle

\thispagestyle{empty}

\begin{abstract}
In this paper, we propose ``SoundSpring'', a cutting-edge error-resilient audio transceiver that marries the robustness benefits of joint source-channel coding (JSCC) while also being compatible with current digital communication systems. Unlike recent deep JSCC transceivers, which learn to directly map audio signals to analog channel-input symbols via neural networks, our SoundSpring adopts the layered architecture that delineates audio compression from digital coded transmission, but it sufficiently exploits the impressive in-context predictive capabilities of large language (foundation) models. Integrated with the casual-order mask learning strategy, our single model operates on the latent feature domain and serve dual-functionalities: as efficient audio compressors at the transmitter and as effective mechanisms for packet loss concealment at the receiver. By jointly optimizing towards both audio compression efficiency and transmission error resiliency, we show that mask-learned language models are indeed powerful contextual predictors, and our dual-functional compression and concealment framework offers fresh perspectives on the application of foundation language models in audio communication. Through extensive experimental evaluations, we establish that SoundSpring apparently outperforms contemporary audio transmission systems in terms of signal fidelity metrics and perceptual quality scores. These new findings not only advocate for the practical deployment of SoundSpring in learning-based audio communication systems but also inspire the development of future audio semantic transceivers.

\end{abstract}



\section{Introduction}\label{section_introduction}

\IEEEPARstart{T}{he} emergence of next-generation wireless networks, particularly 6G, heralds a transformative era in connectivity, ushering in a wide array of applications.
Benefiting from data-oriented signal processing techniques, future wireless networks are expected to not only pursue accurate communication in bit level but also offer a wide range of new functionalities such as semantic-aware intelligent tasks.
Joint signal processing at the transceiver is expected to improve the end-to-end system gain, by sensing and exploiting the intrinsic nature of source signals.
Among these scenarios, audio communication is always the indispensable one.
The tradition transceiver design for audio communication is a divide-and-conquer paradigm.  Audio codecs are meant for compressing audio~\cite{amrwb, opus, AAC}, while the transmission robustness is ensured by channel coding and other error control techniques.
Audio codecs play a pivotal role in compressing the audio, while rate control works in cooperation with the codec to strategically allocate bits across and within audio frames, thereby optimizing the overall communication efficiency.
Channel coding is designed with the pursuit for a low error rate in average. But in practice, we often observe left bit errors that manifest into uncorrectable errors and thus packet loss occurs, in which case the general solution is to request retransmission of lost packets.
However, retransmission is suitable only for scenarios with short round trip times (RTTs). For most real-time communications (RTC) applications, error audio frames may not be concealed via retransmission where we expect audio frames to be played as soon as they are decoded, e.g., FaceTime, WeChat, etc. Resending error audio packets can contribute to overall delay, ultimately leading to a poor user quality of experience (QoE).

An alternative transceiver design paradigm is joint source-channel coding (JSCC)~\cite{fresia2010joint, weng2021semantic, han2022semantic, dai2022nonlinear, xiao2023wireless}, which can well balance the efficiency and robustness over lossy wireless channels with \emph{one-shot} transmission.
However, this approach fuses source signal compression and channel processing completely, which is typically a significant challenge for the deployment of JSCC transceiver in compatible with current layered communication system architecture.
Motivated by that, in this paper, we aim to propose a novel error-resilient audio transceiver which can provide both high transmission efficiency and robust audio communication quality like JSCC with no need of retransmission, while being of the compatible layered architecture.

Here, we discuss the closely related problem of \emph{resilient audio transmission}, that is, designing audio coding methods that can efficiently compress audio signal while also can well mitigate the error propagation caused by packet loss over lossy wireless networks. This problem turns out to be significantly hard to solve, which needs to balance between \emph{efficiency} and \emph{resiliency} since these two terms are naturally contradictory in Shannon information theory. But this is not for a lack of trying, the mainstream solutions can be categorized as the wireless network quality prediction-based forward error correction and the post-processing-based audio signal restoration.

As one broad solution, recent advanced audio codecs that provide high compression efficiency can be combined with forward error correction (FEC) codes to mitigate packet losses~\cite{opus, valin2023low}. However, its efficacy has limited impact on long loss bursts, and inadequate FEC protection may still lead to discordant audio. Inversely, excessive FEC redundancy affects transmission efficiency. The other broad solution is \emph{post} error concealment techniques that restore the impaired audio at the receiver side, such as frame interpolation~\cite{jayant1981effects, amrwb_loss_conceal}. More recently, data-driven packet loss concealment (PLC) solutions have been developed~\cite{contextinpaint, li2022end, msra_plc}, most of which employ a post-processing step on the decoded waveform or spectrum. Consequently, the effectiveness of their concealment is intrinsically linked to the audio codec utilized, limiting their performance with emerging codecs that leverage deep neural networks \cite{valin2019real, kleijn2018wavenet, zhen2020efficient, yao2023variational, soundstream, encodec}. These codecs, characterized by their extensive error propagation range within neural audio decoders, pose challenges to post PLC techniques in providing satisfactory performance. Therefore, a good audio transceiver that balances efficiency and resiliency has still not been found.

In this paper, we aim to propose a pioneering approach to audio transceiver design, emphasizing high compression efficiency while enhancing resiliency against diverse packet loss scenarios. Developing such an audio transceiver poses two fundamental challenges. Firstly, how to explore loss concealment within the \emph{latent feature domain}, diverging from the conventional source domain strategies prevalent in traditional PLC methods? This novel approach aims to improve concealment efficacy and computational efficiency. Secondly, how to investigate strategies to optimize the balance between resiliency and efficiency, with a particular focus on mitigating error propagation during audio decoding? For enhanced compression efficiency, our method involves tokenizing audio latent features utilizing a residual vector quantizer (RVQ), followed by modeling the contextual relationships among these tokens. From the information-theoretic viewpoint, a stronger inter-token dependency correlates with superior compression performance and vice versa. Nonetheless, this increased dependency could extend error propagation in audio decoding. Hence, there exists a critical trade-off between achieving better compression efficiency and ensuring error resiliency.

To tackle aforementioned challenges, as the first work, we propose to integrate the contextual modeling capability of large language models (LLMs) into loss-resilient audio transceiver design. This idea is inspired from the intrinsic consistence between generative modeling and compression, a principle long established in this field \cite{deletang2023language, dai2024deep}.
The impressive \emph{long-range in-context modeling capability} of LLMs brings us fresh insights for designing audio transceivers.
Benefiting from the \emph{mask modelling} mechanism, pre-trained \emph{bi-directional Transformers}, so-called \emph{foundation models} \cite{deletang2023language,bert}, have proven highly successful in a wide range of predictive tasks. Incidentally, the predictive ability of these masked language models (MLMs) can serve both contextual probability estimation in audio compressor and lost token concealment in audio receiver, which vastly simplify error-control in audio communication. Moreover, we adopt the \emph{random masking} strategy for training MLM, which empowers it with the \emph{casual-order context modeling} ability. This simple but efficient strategy can handle both the well-designed specific audio coding dependency pattern and arbitrary unpredictable packet loss patterns. Thus, we find out a new solution to balance efficiency and resiliency.

Specifically, this paper presents \emph{SoundSpring}\footnote{``Spring'' is dubbed here to refer to the ability of a material or object to deform or stretch under force and then return to its original shape once the force is removed. The name SoundSpring is designed with two sides of meaning, indicating the \emph{resilience} property of a \emph{new} type of audio transceiver.}, a novel error-resilient audio transceiver, which has characterizations:

\begin{itemize}
	\item \emph{High compression efficiency}. In addition to the latent feature quantization (RVQ) for obtaining discrete audio tokens, the contextual modeling by MLM contributes to extra compression gain.
	
	\item \emph{Versatility}. SoundSpring supports very flexible multiple-rate audio coding and scalable contextual modeling by using a single model.
	
	\item \emph{Graceful degradation with more lost packets}.
	With the increasing number of lost packets, SoundSpring naturally exhibits graceful quality degradation like that in JSCC.
	
	\item \emph{Real-time transmission}. Real-time audio coding and mask modeling are supported in a streaming variant of SoundSpring for RTC scenarios.
\end{itemize}

We verify the performance gain of SoundSpring on speech and music sources under various packet-loss channel setups, including random packet loss traces as well as wireless local area network (WLAN) packet loss traces. Results demonstrate that, the proposed SoundSpring strikes a commendable balance between efficiency and resiliency against packet loss with no need of retransmission, especially in long burst loss scenarios that was hard to handle by traditional transceivers. In summary, we take an important step towards future LLM-empowered \emph{intelligent and resilient} transceivers, showing the benefit of simplifying error-control mechanism by enabling an LLM to learn diverse contexts among source features.

The remainder of this paper is organized as follows.
 In next section, we briefly review the related work.
 Section \ref{section_method} presents the transceiver design and dual-functional MLM of SoundSpring.
 Section \ref{section_impl} introduces the detailed implementation.
 Section \ref{section_experiment} shows the experiments of our proposed design to illustrate the performance gain, comparing with traditional and neural audio codecs with loss concealment solutions.
 Finally, Section \ref{section_conclusion} concludes the paper.

\textit{Notational Conventions:} Lowercase letters (e.g., $x$) denote scalars and bold ones (e.g., $\x$) denote vectors or matrices. $\x_{i:j}$ denotes a vector slicing from $i$-th element to $j$-th element. $p_x$ denotes a probability mass function with respect to variable $x$. $\mathbb{Z}$ denotes the integer set. $\lfloor \cdot \rfloor$ denotes the floor function.

\section{Related Work}\label{section_related}

\begin{figure*}[t]
	\setlength{\abovecaptionskip}{0.cm}
	\setlength{\belowcaptionskip}{-0.cm}
	\centering
	\includegraphics[width=2\columnwidth]{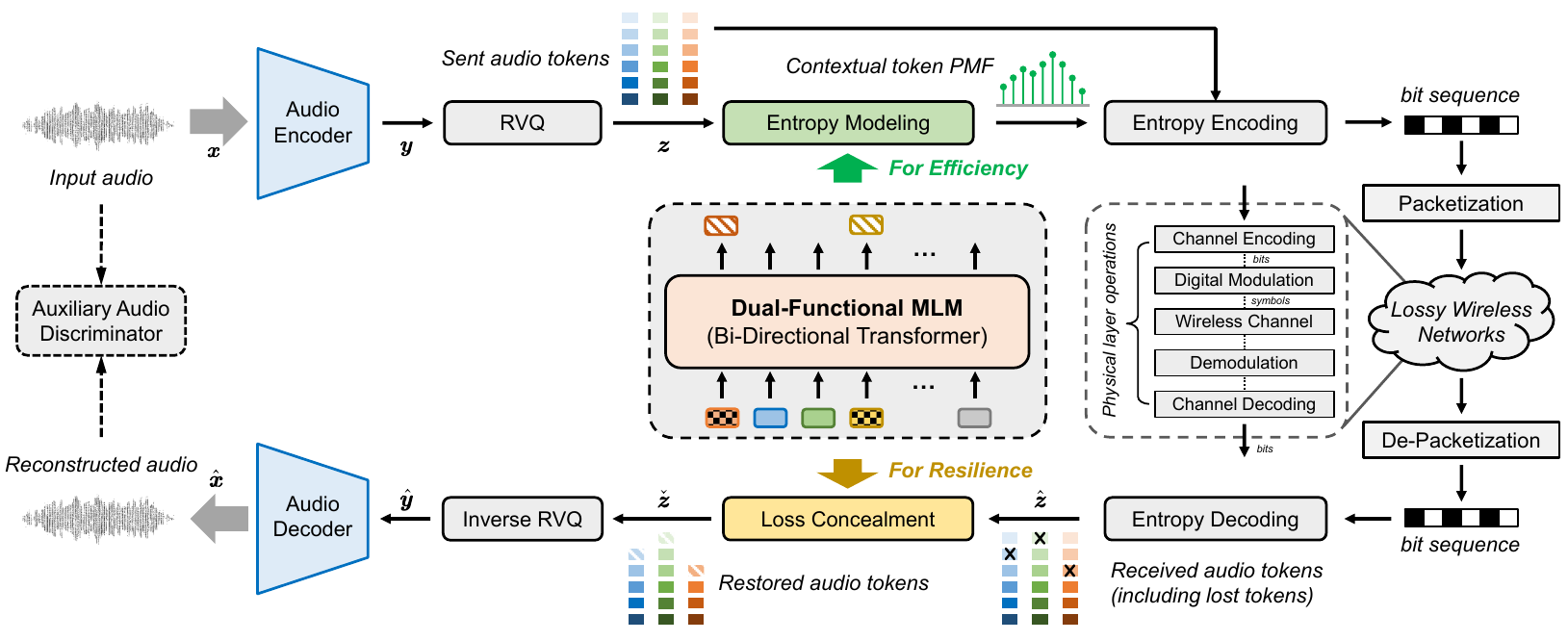}
	\caption{Overview of the architecture of SoundSpring.
		The audio waveform is mapped to audio latent features and then multiple sequences of audio tokens are obtained by residual vector quantizer, which are emitted in order from the sequence in dark color to that in light one.
		A unified dual-functional MLM performs entropy modeling for efficiency at the sender and conceals the lost tokens for resiliency at the receiver. The contextual probability mass
		function (PMF) is leveraged for entropy coding. Given the received tokens, the MLM makes predictions utilizing the neighbouring received frames.
		In both functions, we mask the target tokens and replace them with a mask token.
		In addition, an auxiliary audio discriminator is employed in training to promote audio perceptual quality.
	}\label{fig_overview}
    \vspace{0em}
\end{figure*}

\subsection{Neural Audio Codec}

Audio coding has long been a focal point in communication systems, with recent years witnessing advancements driven by neural networks.
According to the type of signals, neural audio codecs are categorized into acoustic feature codecs~\cite{jiang2023latent, zhen2020efficient, kleijn2018wavenet}, waveform codecs~\cite{soundstream,encodec,yao2023variational}.
Acoustic feature codecs typically rely on a generative model at the receiver, encoding hand-crafted features (e.g. mel spectrogram).
The acoustic feature codecs exploit hand-crafted audio features and rely on powerful audio synthesizers to obtain audio.
In contrast, waveform codecs take in full acoustic information and aims for faithful reconstruction, usually consuming a higher coding rate.
In this work, we consider a waveform codec in SoundSpring.
We reveal that the compression capability of a waveform codec can be potentially further enhanced, and we additionally investigate the resilience of neural audio coding in lossy transmission.

\vspace{-0.06in}
\subsection{Learning-based Audio Transmission and Digital Semantic Communication}

Recent advancements in learning-based audio transmission utilizing JSCC approaches have demonstrated significant robustness against lossy communication channels \cite{weng2021semantic, han2022semantic, xiao2023wireless}.
Wen et al.~\cite{weng2021semantic} developed a semantic transmission framework for speech signals by end-to-end training, mitigating the cliff effect on the audio quality.
Xiao et al.~\cite{xiao2023wireless} proposed a bandwidth-distortion-perception optimizing strategy to strike a better balance between communication efficiency and robustness.
Despite great efforts on designing finite-order symbol constellations to enhance compatibility with existing layered system architectures~\cite{bo2024joint,tung2022deepjscc}, a fundamental challenge of deployment remains because they fuse source signal compression and channel processing completely.
Meanwhile, the robustness and adaptability to varying wireless channels of these systems remain open questions.

Another scheme for digital audio semantic communication is to employ a neural audio codec as mentioned in previous subsection, and maintain the existing physical layer operations. However, despite superior compression performance compared to traditional codecs, the resilience of neural audio codec is still lack of exploration under lossy transmission.
A few works aim to enhance the robustness by introducing adversarial learning~\cite{he2022robust}, or post-processing techniques~\cite{li2022end, msra_plc}.
SoundSpring falls into this category, which has great adaptability to various channel types and we focus on its resiliency in packet-level lossy transmission.

\subsection{Language Modeling}

Language models (LM) have recently made remarkable strides in representation learning~\cite{bert}, text generation~\cite{gpt}, audio generation~\cite{soundstorm, audiogen}, and image and video generation~\cite{yu2023magvit}.
The attention mechanism, a fundamental component in these models, enables strong capabilities in capturing long-term dependencies.
While language models showcase powerful capabilities in audio generation tasks \cite{soundstorm, audiogen}, its application in the area of audio communication systems remains largely unexplored.
Défossez et al.~\cite{encodec} have made the first attempt to use an \emph{auto-regressive} LM to improve compression capabilities, regardless of possible decoding failure in case of transmission errors.
In this paper, we propose a dual-functional language modeling method, which benefits both audio compression and audio concealment.

\vspace{-0.02in}
\section{Method}\label{section_method}

Our goal is to exploit the promising learning capabilities of language model and strike a delicate balance between efficiency and robustness when incorporating compact neural audio representations in audio communication systems.
The general architecture of \emph{SoundSpring} audio transceiver in Fig.~\ref{fig_overview} consists of the following components:

\begin{enumerate}
  \item \emph{A pair of audio encoder and decoder.}
  The audio encoder $\mathcal E$ encodes audio waveform $\x$ into a sequence of compact latent features $\y = \mathcal E(\x) = \{\y_1, \y_2, ..., \y_{T}\}$ with temporal downsampling, followed by quantization.
  At the receiver, the audio decoder $\mathcal D$ reconstructs the waveform $\hat \x = \mathcal D(\hat \y)$ using the reconstructed latent features $\hat \y$.

  \item \emph{Residual vector quantizer (RVQ)} recursively quantizes the residuals of continuous latent features $\y$ into $N_q$ sequences of discrete audio tokens $\z$.

  \item \emph{Dual-functional masked language model (MLM)} is built on top of the discrete audio tokens $\z$ or the impaired received tokens $\hat \z$ for either entropy modeling or loss concealment.

\end{enumerate}

At the transmitter, given the sequence of audio latent features, SoundSpring tokenizes audio latent features using the RVQ.
The trick with RVQ is using a ``stacking'' codebooks, rather than having a single high-resolution codebook, to recursively quantize each $\y_t$ to $z_{t,k}$, $k=1,\cdots,N_q$.
Specifically, we define $N_q$ learnable vector codebook $\E^{(k)}, k=1,2,...,N_q$, having the same dimension as the latent feature.
Latent feature of $t$-th frame $\y_t$ is quantized by selecting the nearest embedding vector from $\E^{(1)}$ (first layer) and the residuals are further quantized by the subsequent VQ layers.
Thus, the latent feature $\y_t$ can be approximated by the sum of the selected vectors $\sum_{k=1}^{N_q}\E^{(k)}_{z_{t,k}}$.

The probabilistic distribution of audio tokens $z_{t,k}$ are estimated by MLM and the entropy encoder accordingly encodes them to bitstream for packetization. Then, data in each packet is channel encoded, modulated to symbols for transmission, like that in conventional communication systems.
Upon receiving, the residual errors which cannot be corrected by channel decoding (error controlling mechanism in the physical layer) will lead to the corruption of related data packet, where the bitstream cannot be decoded by entropy decoder. In case of other types of channel, e.g., packet erasure channels, the state of packet notifies the decoder whether to conceal the audio tokens or not.

In the following subsections, we will start by a general pipeline of MLM on RVQ audio tokens, and then introduce its dual-functionality at the sender and receiver, respectively.

\subsection{MLM on RVQ Tokens}\label{subsection_arch}

\begin{figure}[t]
    \setlength{\abovecaptionskip}{0.cm}
    \setlength{\belowcaptionskip}{-0.cm}
    \centering
    \includegraphics[scale=.48]{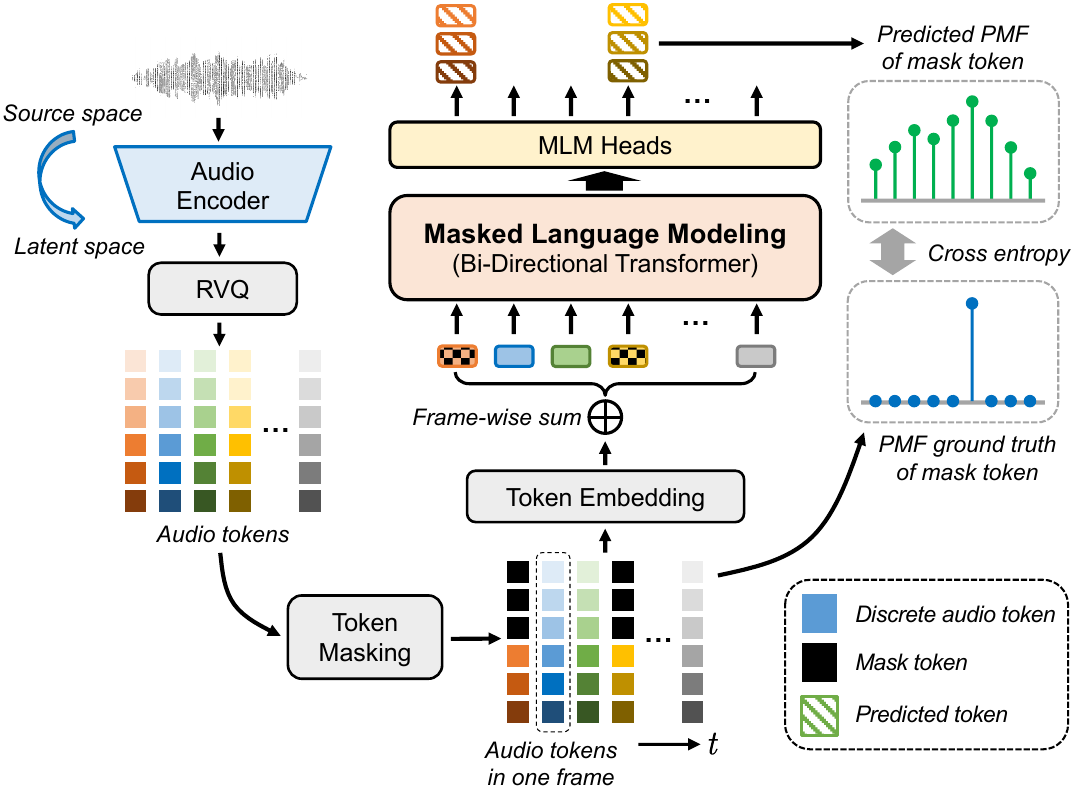}
    \caption{Pipeline of MLM in SoundSpring transceiver. The RVQ audio tokens $\z$ are arranged in a two-dimensional structure, wherein the horizontal orientation denotes temporal axis, while the vertical one is the layer order starting from the bottom in dark color. After token masking, the audio tokens in one frame are projected into embeddings, which are summed and then fed into the bi-directional Transformer.
    The training objective of MLM is to minimize the cross entropy between the ground truth distribution of audio tokens and the predicted PMF of mask token.}
    \label{fig_mlm}
    \vspace{-1em}
\end{figure}

The pipeline of MLM in SoundSpring is illustrated in Fig.~\ref{fig_mlm}.
We adopt a bi-directional Transformer-like~\cite{vaswani2017attention} MLM, denoted as $\Omega$, to model the PMF of the RVQ audio tokens.
Looking along the time axis, we obtain $N_q$ discrete token sequences, with $k$-th sequence $\z^{(k)} = [z_{1,k}, \cdots, z_{T,k}]$ corresponding to $k$-th RVQ layer, $k=1,\cdots,N_q$. Every token is taken from the vocabulary set $\{1,\cdots,M\}$.
We denote $\mathcal M$ as the set of tokens being replaced by the $\mathrm{[MASK]}$ token, with the remaining unmasked ones collected as $\overline{\mathcal M}$.

After token masking, each discrete token $z_{t,k}$ is transformed into an embedding vector $\e_{t,k}$ using a learnable embedding table ($\mathrm{[MASK]}$ token is also mapped to an embedding vector).
The token embeddings of the same frame are summed to obtain $\e_t = \sum_k{\e_{t,k}}$, which is then fed into the Transformer.
Subsequently, $N_q$ MLM heads process the output embeddings, with each head producing the logits of tokens in each RVQ layer.
Therefore, regarding the token $z_{t,k} \in \mathcal M$ which is masked out, we obtain its conditional probability mass function (PMF) as
\begin{equation}
p(z_{t,k} | \overline{\mathcal M}) = H_k \left( \Omega( \{\e_t\}^T_{t=1}; \overline{\mathcal M}) \right),
\end{equation}
where $H_k$ represents the function of $k$-th MLM head, which is a multi-layer perceptron with softmax activation.

Finally, SoundSpring utilizes the PMFs for two functions, one for entropy modeling at the sender to improve efficiency, and the other one for loss concealment at the receiver to improve resiliency.

\subsection{MLM for Audio Compression}\label{subsection_mlm_tx}

While the tokenization delivers a compact discrete representation, the inter-frame dependency of audio tokens is still disregarded.
As Shannon theory says that, the conditional entropy is not larger than the unconditional one, a proper contextual dependency pattern is required to decompose the joint distribution of tokens into a series of contextual token PMFs, thus increasing the coding efficiency.
The auto-regressive factorization as employed in~\cite{vqvae, soundstream} is tailored for audio generation application, with each token conditioned on the past ones.
However, the long-term dependency is susceptible to transmission errors and faces the challenge of increasing computational cost.

To address this challenge, we establish dependency links on groups of tokens, capitalizing on the intrinsic nature of RVQ audio tokens.
RVQ offers versatility, allowing the senders can make coarse-grained rate adjustments by selecting the number of encoded layers or RVQ level $K$, where $K\leq N_q$.
Additionally, it provides scalability, allowing for audio recovery with degraded yet acceptable quality when a few tokens are missing.
However, the scalability results in error propagation along the layer order of tokens, as the tokens are dependent on ones in lower layers.

\begin{figure}[t]
		\setlength{\abovecaptionskip}{0.cm}
	\setlength{\belowcaptionskip}{-0.cm}
	\centering
		\includegraphics[scale=0.56]{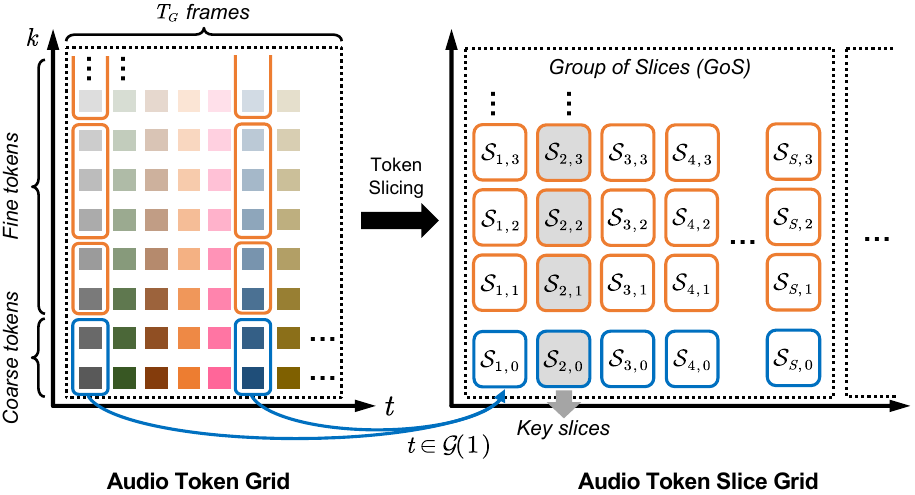}
	\caption{Two-dimensional audio token grid of $z_{t,k}$ and audio token slice grid $\mathcal S_{\ell,j}$.
		Every $T_G$ frames of tokens constitute a group of slices. The audio tokens generated by RVQ in bottom layers are assigned as coarse tokens, while the ones on the top layers are categorized into fine tokens.}
	\label{fig_mlm_compression}
\end{figure}

Considering that the importance of the RVQ tokens along the layer order is monotonically decreasing, we classify the audio tokens into two main categories, as shown in Fig.~\ref{fig_mlm_compression}.
We designate the first $N_C$ RVQ layers of tokens $\z_{t,1:{N_C}}$ as \emph{coarse} tokens (depicted in blue groups), which retain the most fundamental acoustic information.
The remaining tokens in the top $N_F = N_q - N_C$ layers are termed \emph{fine} tokens (depicted in orange groups).
Subsequently, coarse and fine tokens are grouped and each group of tokens $\mathcal S$ is referred to as a \emph{slice}.
On one hand, the first subscript of $\mathcal S$ defines a new axis, named \emph{G-axis}, which is determined by a token slicing strategy $\mathcal G$.
Specifically, $\mathcal G(\ell)$ is the set of timesteps $t$, indicating that the slices $\mathcal S_{\ell, \cdot}$ contain the tokens in $t$-th frame.
To prevent inefficient contextual modeling on long sequences of tokens within the same slice, in which case their correlation is not fully exploited, each $\mathcal G(\ell)$ is restricted to a maximum temporal span.
Thus, we additionally define the concept of Group of Slices (GoS), and every $T_G$ frames constitute a GoS, as illustrated by the dashed box in the figure.
On the other hand, along the layer axis, the coarse tokens of one frame are assigned to a unique slice $\mathcal S_{\ell, 0}$. The fine tokens are assigned to at most $J$ slices in order, depending on the exact number of layers $K$.
Thus, the audio token slice grid in~Fig.~\ref{fig_mlm_compression} can be defined by
\begin{equation}
\mathcal S_{\ell, j} = \bigcup_{t \in \mathcal G(\ell), k \in \mathcal H(j)} z_{t,k},
\end{equation}
where $\mathcal H(j) = \{ N_{j} + 1, \cdots, N_{j+1} \}$,
and $\{N_{j}\}_{j=0}^{J+1}$ is a monotonically increasing sequence, with $N_0=0$, $N_1=N_C$ and $N_{J+1}=N_q$ by definition.

\begin{figure}
		\setlength{\abovecaptionskip}{0.cm}
	\setlength{\belowcaptionskip}{-0.cm}
  \centering
      \includegraphics[scale=1.3]{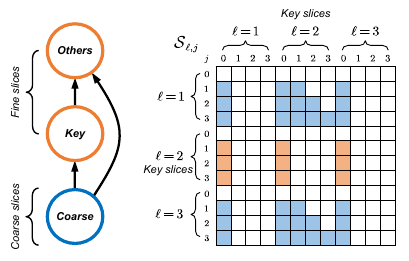}
  \caption{General dependency link of audio tokens and slice-wise dependency pattern $\PPhi$.
  The 2-D token slice grid is flattened into a sequence.
  The dependency of key slices and non-key slices are marked by orange and blue, separately.}
  \label{fig_mlm_compression_dependency}
\end{figure}

Given $\mathcal G(\ell)$, $T_G$, and $\mathcal H(j)$, with respect to each GoS, there are $S$ units along the new G-axis, and at most $J+1$ units along the layer axis.
In addition, we define a special series of slices $\mathcal S_{\ell^{\star}, \cdot}$ as \emph{key slices} per GoS, with $\ell^{\star} \in \{1,\cdots, S\}$.
Under these assumptions, the general dependency link among RVQ audio tokens is formulated, as depicted in Fig.~\ref{fig_mlm_compression_dependency}.
Specifically, it follows the guidelines as below.
\begin{enumerate}
  \item Coarse tokens and fine tokens are placed into separate slices, ensuring that there are no mixed slices.
  \item We establish the entropy model for fine tokens only. Coarse tokens are crucial for maintaining audio fidelity and consume a small portion of coding rate.
      We do not venture to acquire the coding gain on coarse tokens.
  \item Tokens within the same slice are modeled by the MLM in parallel, which are independent of each other. They are encoded and then collected into one packet. We avoid modeling the intra-slice dependency to simplify and accelerate entropy modeling.
  \item The fine tokens in key slices are exclusively conditioned on the coarse token slices within the GoS, while the ones in non-key slices are additionally conditioned on the key slices.
      Specifically, the slice $\mathcal S_{\ell, j}, j>0$ is encoded based on key slices $\mathcal S_{\ell^{\star}, 1:j}$ as well as the coarse tokens. This is consistent with the causality of RVQ tokens and consistent with the token embedding summation for each frame.
\end{enumerate}

For better visualization, we summarize the dependence using a binary dependency matrix $\PPhi$, as displayed in Fig.~\ref{fig_mlm_compression_dependency}.
Each slice (each row of the matrix) is conditioned on the slices which are filled in color, while being independent of others.  For example, we want to model the PMF of $\mathcal S_{3,3}$, with corresponding key slices $\mathcal S_{2,1:3}$.
The MLM initiates by modeling its key slices $\mathcal S_{2, 1:3}$. It firstly uncovers the coarse tokens in the same GoS, $\overline{\mathcal M} = \mathcal S_{1:S,0}$, and masks the other tokens.
The MLM takes as input the summation of $N_4$ layers of embeddings.
We obtain the PMF of key slices $p_{\z}(\mathcal S_{2, 1:3} \vert \overline{\mathcal M})$.
Then, the logits of the target slice are computed to retrieve the PMF of tokens $p_{\z}(\mathcal S_{3, 3} \vert \overline{\mathcal M})$ with $\overline{\mathcal M} = \mathcal S_{1:S,0} \cup \mathcal S_{2, 1:3}$.

\begin{figure}[t]
	\setlength{\abovecaptionskip}{0.cm}
	\setlength{\belowcaptionskip}{-0.cm}
	\centering
	\includegraphics[width=0.85\columnwidth]{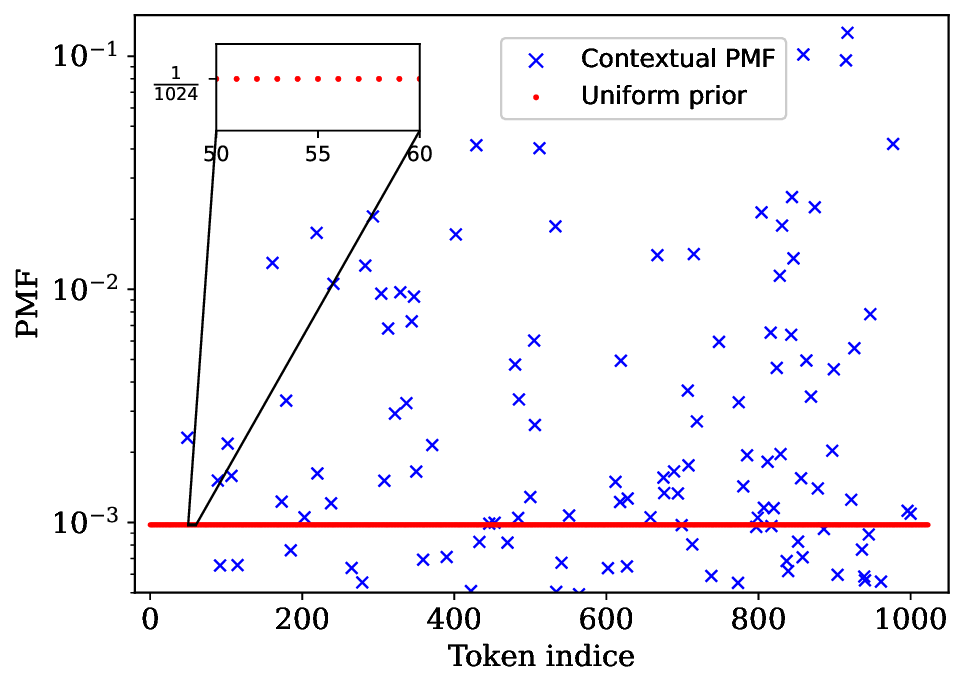}
	\caption{PMF of an exemplar audio token in the first fine layer, $M=1024$. Compared to a uniform prior distribution, the contextual PMF is quite unbalanced and sparse, which leads to the reduction in coding rate.}
	\label{fig_pmf_visualization}
    \vspace{-.1em}
\end{figure}

In Fig.~\ref{fig_pmf_visualization}, we present a visualization demo of the PMF of an exemplar audio token in layer $k=N_c+1$.
Without entropy modeling, the codelength required to encode any $z_{t,k}$ is $\log_2M=10$ bits, with uniform prior assumption.
Instead, the contextual token PMF output by MLM is unbalanced and sparse across the vocabulary of the codebook. The receiver could losslessly recover $z$, provided that its \emph{condition} tokens on which its PMF is conditioned at the sender, are correctly received. In this case, only $3.83$ bits are required.

\textbf{Objective.} For the sender, the MLM only adapts to an agreed coding dependency pattern $\PPhi$ between the sender and the receiver, and thus decreases the averaged codelength required to encode fine tokens.
It is accomplished by minimizing the cross entropy between the PMF of ground truth of mask tokens $q(\z)$ and the contextual PMF $p(\z \vert \overline{\mathcal M};\PPhi)$, which is the expected length of the coded datastream of mask tokens.

\begin{figure*}
		\setlength{\abovecaptionskip}{0.cm}
	\setlength{\belowcaptionskip}{-0.cm}
  \centering
  \includegraphics[scale=1]{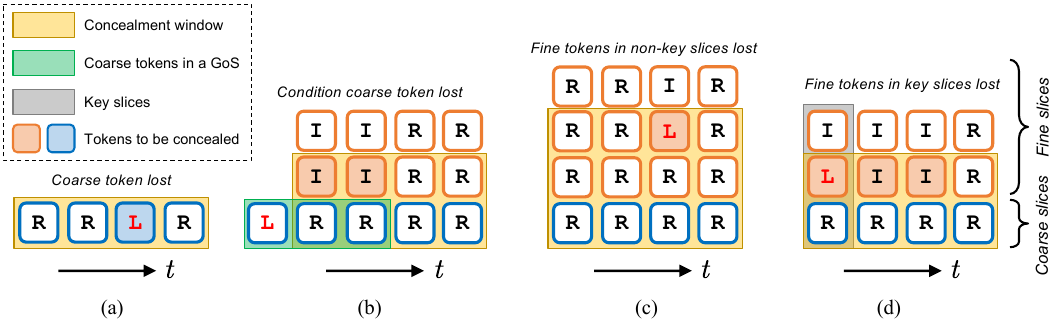}
  \caption{Possible received token grid patterns consisting of $\textbf{\texttt{R}}$, $\textbf{\texttt{L}}$ and $\textbf{\texttt{I}}$ tokens. For each pattern, we present the target tokens to be concealed (filled in color), and the composition of concealment windows (yellow boxes). A single unit in the grid is the surrogate for multiple tokens for concise illustration.}
  \label{fig_token_lost_pattern}
\end{figure*}

\begin{figure}
		\setlength{\abovecaptionskip}{0.cm}
	\setlength{\belowcaptionskip}{-0.cm}
  \centering
  \includegraphics[scale=1.2]{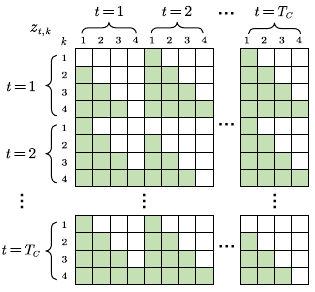}
  \caption{Concealing dependency matrix $\PPsi$. The dependency is bi-directional along the time axis and causal along the layer axis. In practical, the masking pattern is obtained according to the state of tokens in the concealment window.
  }
  \label{fig_mlm_concealment}
\end{figure}

\subsection{MLM for Loss Concealment}\label{subsection_mlm_rx}

SoundSpring's receiver needs to accomplish the following tasks to recover audio.
\begin{itemize}
  \item Collect context referring to dependency pattern $\PPhi$ and entropy decoding.
  \item Token prediction for loss concealment.
  \item Recover audio latent features by inverse RVQ, and then restore the audio waveform.
\end{itemize}

To decode fine tokens, we will utilize the exact coding dependency pattern $\PPhi$ established in subsection~\ref{subsection_mlm_tx} to obtain PMF from MLM, which is required for entropy decoding.
After entropy decoding, there exist three types of tokens: received tokens (\textbf{\texttt{R}}), lost tokens (\textbf{\texttt{L}}), and invalid tokens (\textbf{\texttt{I}}).
The invalid token is possibly attributed to two causes. The lost of its condition tokens leaves it labeled by \textbf{\texttt{I}}. In addition, due to the recursive nature of RVQ, the tokens will be invalid if any slice in lower levels in the slice grid is not correctly recovered.
In practice, based on the coding dependency $\PPhi$, there may exist four cases as depicted in Fig.~\ref{fig_token_lost_pattern}.
To conceal lost tokens, we also restrict a \emph{concealment window} (the yellow boxes in the figure), containing at most continuous $T_C$ frames.

\subsubsection{Case 1} On condition that coarse tokens are lost, the MLM masks these lost coarse tokens while uncovering other $\textbf{\texttt{R}}$ coarse tokens. Only $N_C$ layers of coarse tokens are projected into embeddings, which are summed as input of MLM.
Distinguished from the sender, where the correlation among coarse tokens is disregarded, the MLM at the receiver leverages the neighbouring coarse tokens to predict the missing ones.

\subsubsection{Case 2} It is the chain effect of Case 1. Despite that all the coarse tokens in the concealment window are received, there exist cases where condition coarse tokens outside the window are lost. It is possible when the GoS (green box) is not exactly coincident with concealment window, and when the GoS is not continuously distributed in temporal domain.
In these cases, all the fine tokens in the associated frames are labeled as \textbf{\texttt{I}}. The MLM conceals these $\textbf{\texttt{I}}$ tokens in the first few fine layers provided that the coarse ones in the concealment window are $\textbf{\texttt{R}}$ tokens. Empirical observations suggest that predicting all the fine tokens is not as effective as predicting the first few layers of tokens due to the error propagation along the layer order.

\subsubsection{Case 3} Coarse tokens are all received but fine tokens in non-key slices are lost.
To conceal $\textbf{\texttt{L}}$ token $z_{t,k}$ at $k$-th layer, $k$ layers of tokens are included for concealment. Tokens above $k$-th layer $\z_{t,>k}$ in the same frame are all labeled as $\textbf{\texttt{I}}$ and will not be concealed, arising from the causality of RVQ.

\subsubsection{Case 4} When the fine tokens in key slices are lost, the associated tokens in non-key slices are invalid (second and third column of the grid).
On condition that the tokens in lower layers are $\textbf{\texttt{R}}$ tokens, these invalid tokens will be included in the concealment window, similar to Case 2.

It is worth noting that not all missing tokens are to predicted. Within a frame, when token errors occur in multiple layers, the ones at lower levels are predicted, while the prediction at high levels is found not to yield positive impact on audio quality enhancement.
Pragmatically, the mask modeling of MLM follows the concealing dependency $\PPsi$, as shown in Fig.~\ref{fig_mlm_concealment}.
According to actual situations of decoded tokens, the masking pattern is obtained from $\PPsi$ for the tokens in the concealment window, where the tokens to be concealed are replaced by $\mathrm{[MASK]}$ tokens.
After summing up the embeddings of each frame, the MLM computes the PMF and applies the maximum-likelihood criterion to make predictions for the missing tokens.

\begin{figure}[t]
		\setlength{\abovecaptionskip}{0.cm}
	\setlength{\belowcaptionskip}{-0.cm}
  \centering
  \includegraphics[width=0.8\columnwidth]{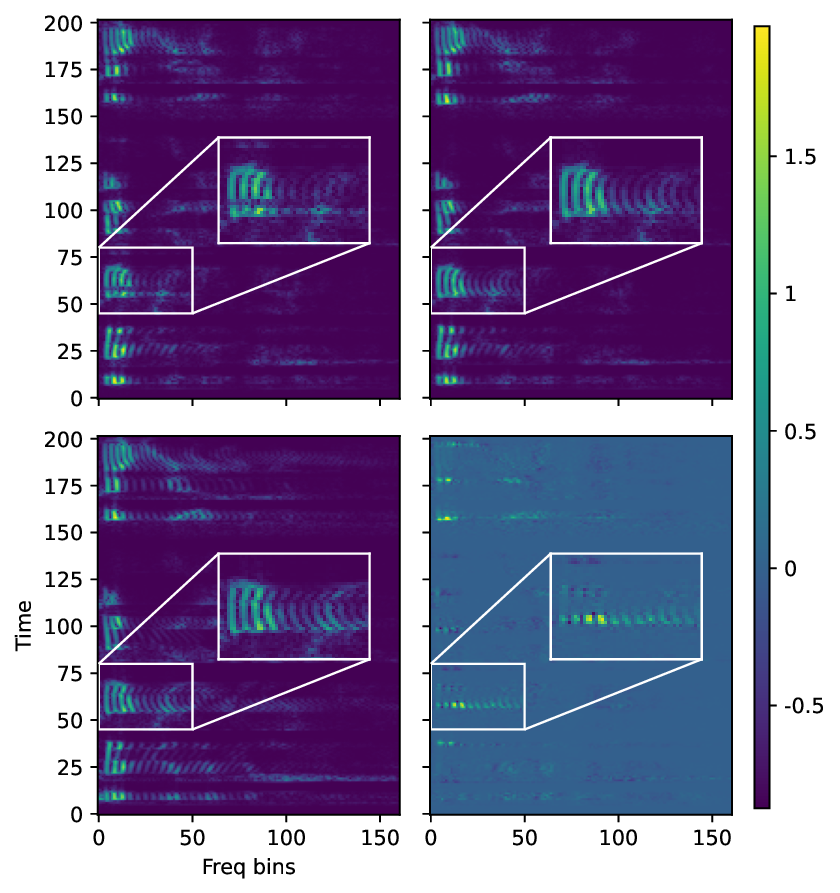}
  \caption{Spectrum visualization demo of the spectrum of reconstructed audio using the received token $\hat{\z}$ (top left), and using the concealed tokens $\check{\z}$ by MLM (top right). The details in the box are enlarged for better visualization. The bottom figures show the spectrum of the original audio (bottom left) and the differences arising from the concealment (bottom right).}\label{fig_spec_visualization}
\end{figure}

A toy example illustrating the effects of loss concealment is displayed in Fig. \ref{fig_spec_visualization}.
It can be observed that the gap arising from the packet loss is seamlessly filled in, and the power spectrum of the concealed audio closely resembles that of the raw audio (bottom left subfigure).

\textbf{Objective.} The objective of MLM for loss concealment is maximizing the prediction accuracy when tokens are lost due to transmission. This is achieved by optimizing a classification problem to minimize the cross-entropy between the PMF of ground truth of missing tokens and the MLM predicted PMF of the lost or invalid tokens.
Due to the unpredictable nature of channel errors, the MLM must adapt to various dependency patterns as discussed above, where the $\textbf{\texttt{R}}$ tokens in the concealment window constitute the unmasked token set $\overline{\mathcal M}$.
Incidentally, this objective function is identical in format as that in audio compression and the loss function in training can be unified as a log-loss over the mask tokens, with condition tokens as input.
The distinction is that the contextual pattern $\PPhi$ for entropy modeling is predefined by the transceiver, while the dependency pattern $\PPsi$ for loss concealment is diverse, depending on the received tokens.
Thus, we design a specific masking strategy in training to adapt SoundSpring to these admissible patterns in Fig.~\ref{fig_token_lost_pattern}, which will be introduced in the following section.

\subsection{The Trade-off between Efficiency and Resiliency} \label{subsection_tradeoff}

SoundSpring naturally exists a trade-off between efficiency and resiliency when designing the contextual dependency pattern.
Strong contextual relationships can effectively fit the real distribution of audio tokens, leading to better compression gain.
However, it also has negative impacts. The complex dependencies leave them vulnerable to channel transmission errors, leading to large-scale error propagation.
That is the motivation why we choose not to model coarse tokens contextual relations for further compression.
It is verified from the experiments that the impact of losing concise tokens outweighs the compression gains from modeling their contextual relationships.
Instead, FEC can be introduced to enhance the robustness of coarse tokens, which is discussed in the following section.

Additionally, inter-slice dependency of fine tokens is also not fully exploited at the sender.
For example, in Fig.~\ref{fig_mlm_compression_dependency}, $\mathcal S_{3,3}$ does not rely on $\mathcal S_{3,2}$.
The joint PMF of $\mathcal S_{\ell,v=1:j}$ are decomposed to the product of PMFs of slices.
It is basically out of the concerns about error propagation due to complex dependencies, and concerns about increasing time complexity.
Conversely, at the receiver, we aim to leverage as many tokens as possible to conceal the missing tokens.
Despite that we can optimize the dual-functional MLM towards superior efficiency by overfitting a unique coding dependency $\PPhi$, the MLM still needs to be adaptive to the diverse concealing dependencies, because the channel transmission error is random.
Thus, it is possible to achieve a balance between efficiency and resiliency by adapting MLM to various dependency patterns.
Besides, the trade-off problem is also constrained by other potential factors, e.g., causality, real-time factor.

\section{Implementation of SoundSpring} \label{section_impl}

In this section, we introduce the implementation of SoundSpring, including the integration of forward error correction (FEC) codes, token slicing strategies, modular implementation details, and finally the training recipe of modules in SoundSpring.

\vspace{-.1in}
\subsection{SoundSpring with FEC} \label{subsection_fec}

While SoundSpring endeavors to conceal coarse tokens in the event of transmission error, the cost of losing these foundation tokens is still considerable, given their role as conditions of fine tokens.
Apart from adding more redundancy by channel coding before transmission, injecting redundancy in source coding is also feasible to enhance robustness.
We attempt to impose addition redundancy for protection of coarse tokens against lossy transmission.
Experimental results show that FEC of coarse tokens greatly ensures fundamental quality with minimal cost of bandwidth efficiency.
However, it is crucial to note that, the FEC solution has its own limitation. In the presence of long burst losses, the backup codes may also be lost, while reducing the transmission efficiency.
Nevertheless, we will demonstrate that through joint efforts with FEC, SoundSpring is proved to effectively enhance the resilience, where MLM is well generalized to various channel conditions.

\subsection{Token Slicing Strategy}

In this subsection, we discuss the token slicing strategy $\mathcal G$, i.e., how to partition the tokens into slices, in two typical scenarios.

\subsubsection{Non-streaming scenarios}

In general audio transmission scenarios, e.g., transferring audio files on social media, SoundSpring can effectively utilize the full context to optimize the coding efficiency, without the demand of causal processing.

Since the tokens within the same slice are independent of each other, we choose to maximize the inter-slice dependency, such as to obtain the coding gain.
Considering the fact that the tokens are mostly correlated with those in the neighbouring frames, a periodical $\mathcal G(\ell)$ in temporal domain could keep distance among tokens in the same slice, while fully exploiting the slice-wise dependency.
Specifically, the periodical token slicing for one GoS with $S$ units is formulated by
\begin{equation}\label{eq_period_slice}
  \mathcal G(\ell) = \left\{t \vert t = Sn + \ell,  n \in \mathbb Z, 0 < t \leq T_G \right\},
\end{equation}
with $ \ell = 1,\cdots,S$.

\subsubsection{Streaming scenarios}

In real-time streaming applications, such as Facetime call over VoIP, online conferencing, there are challenges related to latency and the need for strict adherence to temporal causality.
The token slicing strategy $\mathcal G$ as in~\eqref{eq_period_slice} is infeasible and inevitably causes a long latency.
Following the Real-time Transport Protocol (RTP) protocol \cite{RFC7587}, where a few frames are grouped into an RTP packet, in SoundSpring, tokens from each frame are collected into a single slice.
For simplicity, SoundSpring assigns each frame a unique slice index, denoted by $\mathcal G(\ell) = \ell, \ell=1,2,\cdots,T$.

\begin{figure}[t]
		\setlength{\abovecaptionskip}{0.cm}
	\setlength{\belowcaptionskip}{-0.cm}
	\centering
	\includegraphics[width=\columnwidth]{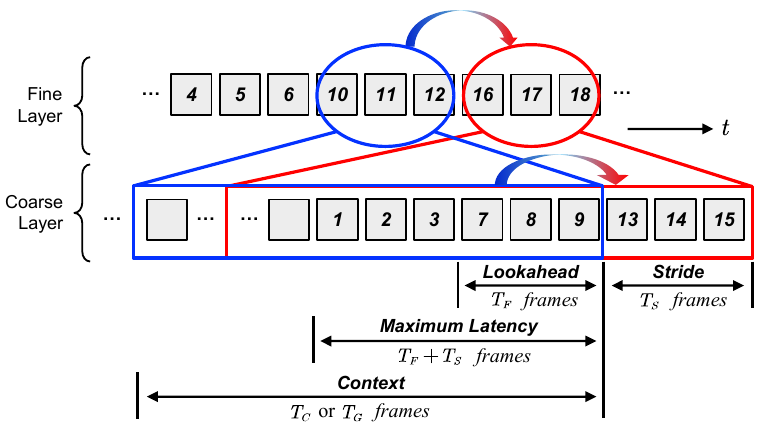}
	\caption{Exemplar diagram of audio token processing for dual-functional MLM in streaming SoundSpring-S.
		The GoS or concealment window slides with a stride along the time axis, producing PMF estimation or token predictions for each inference. The number of contextual frames for entropy modeling and loss concealment is $T_G$ and $T_C$ frames, respectively, which is consistent with previous definitions.
		The stride is $T_S=3$ frames, and the number of lookahead frames is $T_F=3$ in this example.
		The digits in the boxes indicate the encoding and concealing order of tokens.
		For instance, to encode the fine token labeled `10' to `12', their PMFs require the conditioning token labeled `7' to `9', resulting in a maximum latency of 6 frames.}\label{fig_streaming}
\vspace{-1.2em}
\end{figure}

Fig.~\ref{fig_streaming} displays the streaming processing schedules of the MLM.
Each inference involves $T_G$ frames for entropy modeling and $T_C$ frames for loss concealment, as defined previously.
In contrast to the non-streaming inference, the GoS slides with a stride of $T_S$ frames, producing the PMF estimation for $T_S$ frames.
In addition, we also allow few lookahead frames in encoding and concealment, introducing an initial latency of $T_F$ frames.
It is completely causal without initial latency when $T_F$ is zero.
Thus, the overall maximum latency is $T_S + T_F$ frames.
Considering that the audio tokens are streaming faster than frame-wise inference of MLM, we choose stride of $T_S > 1$.
We name the SoundSpring in streaming applications as \emph{SoundSpring-S}.
In section~\ref{section_experiment}, we discuss the compromise on performance for SoundSpring-S to meet the causality and latency requirement.

\subsection{Modular Implementation Details}

Following~\cite{soundstream}, we adopt a stack of 1-D residual convolutional neural networks with multiple kernel sizes as the pair of audio encoder $\mathcal E$ and decoder $\mathcal D$, in order to capture multi-resolution audio features. We employ four convolutional blocks having strides of $(2, 4, 5, 8)$, yielding 320-fold temporal downsampling.
In streaming scenarios, i.e., SoundSpring-S, all operations in the audio encoder $\mathcal E$ and decoder $\mathcal D$ are strictly causal, including the convolution operators.

For SoundSpring, the MLM consists of 24 Transformer blocks with dimension $d=768$.
For SoundSpring-S, these parameters are set to 12, and 512, respectively, in order to reduce the computation cost.
Each codebook in multi-layer RVQ has $M=1024$ entries.
For 16 kHz mono audio input, the number of RVQ layers is set as $N_q=36$, with $N_C=3$ layers of coarse tokens, while for 48 kHz stereo audio, the number of RVQ layers is $N_q=16$, and each layer of uncoded tokens accounts for 1.5 kbps bitrate, with $N_C=2$ layers of coarse tokens.
$M+1$ tokens including the $\mathrm{[MASK]}$ token are projected to $d$-dimensional embedding features, and then fed to the MLM.

\subsection{Training Recipe}\label{subsection_train}

The training recipe of SoundSpring can be generally divided into two stages: training the neural audio codec and the dual-functional masked language model.

\subsubsection{Training of audio codec}

The optimization of the neural audio codec with $N_q$ layers of RVQ quantizer follows the practice of~\cite{encodec}, which combines a reconstruction loss, VQ commitment loss $L_{\mathrm{VQ}}$, plus an adversarial loss.
The VQ commitment loss $L_{\mathrm{VQ}}$ is calculated by the distance between the latent feature and its quantized value counting the gradients of input feature only. The gradient of the audio encoder is obtained through a straight-through estimator~\cite{bengio2013estimating}, while the codebook update uses exponential moving averages (EMA).

The reconstruction loss is defined in waveform domain $L_{\mathrm{wav}} = \Vert \x-\hat \x \Vert_{1}$, and the frequency domain $L_{\mathrm{freq}}$, which combines a 1-norm loss and a 2-norm loss calculated over the mel spectrogram, i.e.,
\begin{equation}\label{eq_loss_freq}
  L_{\mathrm{freq}} = \sum_{\x} {\Vert M_{\x} - M_{\hat \x} \Vert_1 + \Vert \log M_{\x} - \log M_{\hat \x} \Vert_2},
\end{equation}
where $M_{\x}$ and $M_{\hat \x}$ denote the mel spectrogram of the raw and reconstructed audio.

Generative adversarial networks (GANs), one of the most dominant generative models, have been widely applied to audio synthesis~\cite{melgan}. The discriminator estimates the probability that a sample comes from the real data distribution $p(\x)$. Regarding the audio decoder $\mathcal D$ as the generator, an additional discriminator network is jointly optimized together with the audio encoder and decoder. The training strategy, which is often called adversarial training, aims to maximize the probability that the generated waveform comes from $p(\x)$ when optimizing the audio decoder, while minimizing it when optimizing the discriminator.
We employ a multi-scale discriminator $\tilde{\mathcal D}$ as in~\cite{melgan}.
The adversarial loss for the audio codec is formulated as $L_{\mathrm{adv}} = \frac{1}{U}\sum_{u}{\max\left(0, 1 - \tilde{\mathcal D}^U_{u=1}\left(\hat \x \right)\right)}$, where $U$ is the number of scales, together with a feature-matching loss $L_{\mathrm{feat}}$, which is the 1-norm distance between the intermediate features of $\tilde{\mathcal D}$ obtained from $\x$ and $\hat{\x}$.
Therefore, the overall loss for training the neural audio codec is
\begin{equation}\label{eq_loss_codec}
  L = \lambda_t L_{\mathrm{wav}} + \lambda_f L_{\mathrm{freq}} + \lambda_{\mathrm{adv}} L_{\mathrm{adv}} + \lambda_{\mathrm{feat}} L_{\mathrm{feat}} + L_{\mathrm{VQ}},
\end{equation}
where $\lambda_t, \lambda_f, \lambda_{\mathrm{adv}}, \lambda_{\mathrm{feat}}$ are hyperparameters to balance the loss terms.
In practice, the multi-scale discriminator has $U=5$ scales to capture features with different resolutions from 128ms to 8ms length of audio segment, with different STFT window lengths.
During training, we sample RVQ level $K$ uniformly from $\{N_C,\cdots,N_q\}$ to adapt the audio codec to versatile transmission.

\subsubsection{Training of dual-functional MLM}

The proposed dual-functional MLM distinguishes fundamentally from the language model for representation learning~\cite{bert}, which typically employs a fixed masking ratio. The random packet loss necessitates the adaptation of the MLM to diverse dependency patterns. The versatility and adaptivity sets SoundSpring MLM from existing language models.
Therefore, we design the masking schedule of training SoundSpring MLM from the following two perspectives.

Firstly, given a masking scheduling function $\beta(\tau) \in (0,1)$, which defines the masking ratio of frames, and $\tau$ is randomly sampled from a uniform distribution $\tau \sim \mathcal{U}(0,1)$.
We will mask $\lfloor T\beta(\tau) \rfloor$ frames along the temporal domain.
Specifically, we use $\beta(\tau) = \frac{1}{2}\left(1 + \cos(\tau\pi)\right)$, which leans towards masking a large proportion of frames for entropy modeling, or masking a small ratio of frames in alignment with loss concealment.

Secondly, RVQ level $K$ is uniformly from $\{N_C,\cdots,N_q\}$ to simulate encoding $K$ layers of tokens for transmission.
Besides, another level $k$ is also randomly sampled from $\{1,\cdots,K\}$.
The tokens in those $\lfloor T\beta(\tau) \rfloor$ frames are masked from the $k$-th layer to the $K$-th layer, while those in other frames are all unmasked.
This enables the adaptation to multiple coding rates as well as dependencies, including both coding dependency $\PPhi$ and concealing dependency $\PPsi$.

Following this masking strategy, the dual-functional MLM is trained using the unified log-loss function on the mask tokens only, ensuring that it adapts to versatile audio transmission and adapts to diverse states of received tokens.

\section{Experimental Results}\label{section_experiment}

In this section, we provide experimental results (including subjective user ratings) to quantify the gains of our proposed SoundSpring versus conventional coded transmission methods.

\subsection{Datasets and Experimental Settings}\label{subsection_exp_settings}

\subsubsection{Datasets}

The audio data used in training is based on two datasets: LibriSpeech dataset~\cite{librispeech} and MTG-Jamendo dataset~\cite{mtgjamendo}.
The former is a collection of speech segments from audiobooks, where the speech signals are sampled at 16 kHz. The latter is a stereo music dataset, which contains over 55,000 tracks with diverse genres and artists.
The music are all resampled to 48 kHz, for convenience.
The artists of the audio tracks are not overlapped between the training and the test set.

\subsubsection{Training setup}

We train the audio codec and MLM using the Adam optimizer~\cite{kingma2014adam} with learning rate as $5\times 10^{-5}$, and the mini-batch size is set as 64. The learning rate of training the audio codec and audio discriminator is $3\times 10^{-4}$ and $10^{-4}$. Following the training strategy in~\cite{encodec}, the weights of loss terms $\lambda_t, \lambda_f, \lambda_{\mathrm{feat}}, \lambda_{\mathrm{adv}}$ are set as $0.1, 1, 2, 2$, separately. In practice, we initialize the parameters of audio codec using the open-source checkpoint and finetune it in a few epochs on our dataset.

\subsubsection{Packet-loss channel setup}

We conduct our simulations on memoryless packet lossy channels with a given loss ratio as well as a WLAN channel.
The packet loss trace of WLAN channel is simulated by a three-state Markov model~\cite{milner2004}, where the average packet loss ratio is around 13\%.

\begin{figure*}[t]
	\setlength{\abovecaptionskip}{0.cm}
	\setlength{\belowcaptionskip}{-0.cm}
	\centering
	\includegraphics[scale=0.43]{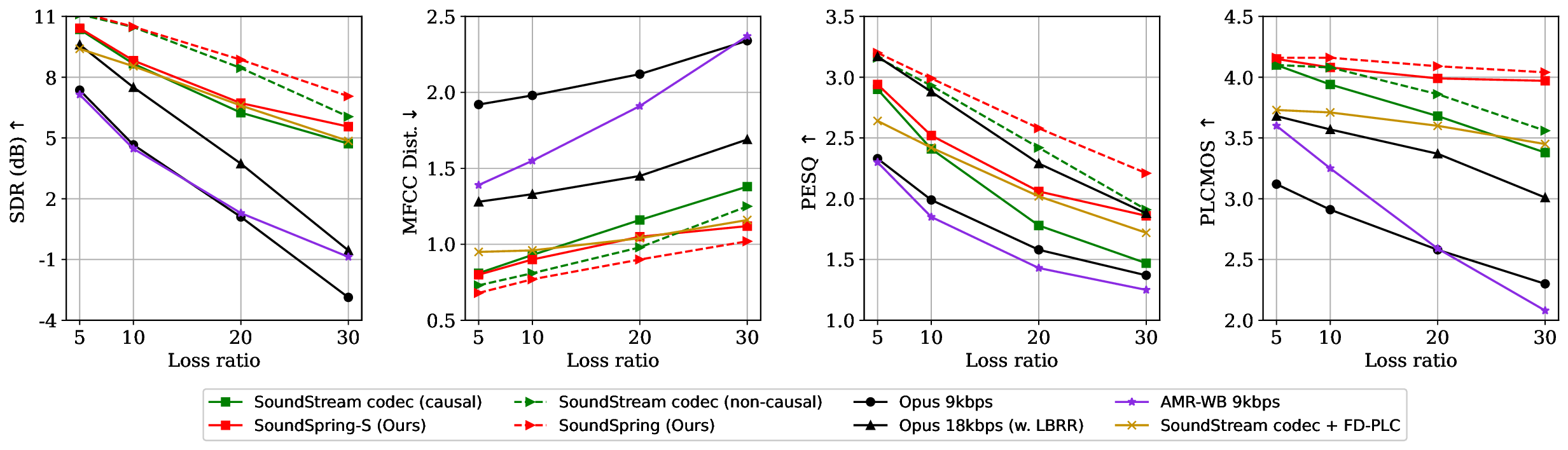}
	\caption{Quality versus loss ratio which are reported in objective distortion metrics and subjective quality scores over channels with i.i.d. random packet loss patterns. The audio is encoded in 5.0 kbps except Opus and AMR-WB. SoundSpring-S is configured with the group size of $T_S=3$ frames, $T_C=T_G=20$ contextual frames, and $T_F=2$ lookahead frames. The upward arrow $\uparrow$ indicates that higher values of the metric are favorable, and vice versa.}
    \label{fig_dmr}
    \vspace{0em}
\end{figure*}

\subsubsection{Baseline methods}

Adaptive multi-rate wideband (AMR-WB)~\cite{amrwb} and Opus~\cite{opus} are provided as the representatives of traditional audio codecs. Bitrate of AMR-WB ranges from 6 kbps through 24 kbps, yet with predetermined options.
Opus is the state-of-the-art lossy audio coding format that has been widely used in VoIP applications.
In the following experiments, we use Opus for both speech coding and music coding.
Advanced audio coding (AAC)~\cite{AAC} is another widely-used audio codec used for high-quality audio delivery over the Internet, which supports a wider range of bitrates.
The loss of transmitted packets is compensated by post-processing techniques in these codecs.
AAC adopts frame interpolation as its loss concealment method.
Aside from frame interpolation for loss concealment, Opus also allows embedding in-band FEC codes, a.k.a., LBRR (low bit-rate redundancy). As defined by Opus protocol, LBRR is mandatorily disabled when the coding bitrate is low.
Therefore, it is fair to compare these engineered codecs with our proposed SoundSpring, which is a framework that combines the neural audio coding and loss concealment.
For SoundSpring, the concealment takes effect only on the audio tokens without any post-processing operations on audio waveform.

\begin{table}[h]
	\renewcommand{\arraystretch}{1.3}
	\centering
	\normalsize
	\caption{Efficiency and resiliency performance evaluated over 48kHz music data.}
	
	\begin{adjustbox}{width=\columnwidth,center}
		\begin{tabular}{!{\vrule width1pt}m{.04\columnwidth}|m{.08\columnwidth}|m{.14\columnwidth}|m{.12\columnwidth}!{\vrule width1pt}m{.14\columnwidth}|m{.14\columnwidth}|m{.16\columnwidth}!{\vrule width1pt}}
			
			\Xhline{1pt}
			
			\centering $K$ & \centering MLM & \centering \makecell[c]{Bitrate \\ (kbps)} & \centering Packet Loss (\%) & \centering \makecell[c]{SI-SNR \\ (dB) $\uparrow$} & \centering \makecell[c]{SDR \\ (dB) $\uparrow$} & \centering ViSQOL $\uparrow$ \tabularnewline
			
			\Xhline{1pt}
			
			\centering \multirow{2}*{4} & \centering \ding{55} & \centering 8.9  & \centering \multirow{2}*{10} &
			\centering 5.06 & \centering 5.72 & \centering 3.81   \tabularnewline
			
			\cline{2-3}\cline{5-7}
			
			~  & \centering \ding{51} & \centering \textbf{7.6}  & ~ & \centering \textbf{6.27} & \centering \textbf{7.21}  & \centering \textbf{3.90}   \tabularnewline
			
			\hline
			
			\centering \multirow{3}*{8} & \centering \ding{55} & \centering 14.9  & \centering \multirow{2}*{10} & \centering 5.08 & \centering 5.87 & \centering 3.83 \tabularnewline
			
			\cline{2-3}\cline{5-7}
			
			~ & \centering \ding{51} & \centering \textbf{12.2} & ~ & \centering \textbf{6.72} & \centering \textbf{7.55} & \centering \textbf{3.93} \tabularnewline
			
			\cline{2-7}
			
			~ & \centering - & \centering 14.9  & \centering 0  & \centering 10.38 & \centering 11.24 & \centering 4.11 \tabularnewline
			
			\hline
			
			\multicolumn{2}{!{\vrule width1pt}c|}{\multirow{3}*{Opus}} & \centering \multirow{2}*{9.0} & \centering 5 & \centering 1.86 & \centering 3.37 & \centering 3.35 \tabularnewline
			
			\cline{4-7}
			
			\multicolumn{2}{!{\vrule width1pt}c|}{~} & & \centering 10 & \centering -1.26 & \centering 1.12 & \centering 3.30 \tabularnewline
			
			\cline{3-7}
			
			\multicolumn{2}{!{\vrule width1pt}c|}{~} & \centering 18.0 & \centering 10 & \centering 2.71 & \centering 4.37 & \centering 4.07 \tabularnewline
			
			\hline
			
			\multicolumn{2}{!{\vrule width1pt}c|}{AAC} & \centering 112 & \centering 10 & \centering 4.58 & \centering 4.74 & \centering 3.88 \tabularnewline
			
			\Xhline{1pt}
			
		\end{tabular}
	\end{adjustbox}
	\label{tab_music}
\end{table}

We provide the baseline performance of SoundStream codec~\cite{soundstream}, which uses the same neural audio codec and RVQ quantizer, except the dual-functional MLM. We also compare SoundSpring audio transceiver with a data-driven packet loss concealment method in feature-domain (FD-PLC)~\cite{msra_plc} using SoundStream codec.
In particular, FD-PLC method aims to improve the robustness of neural audio coding by joint training with decoder, which sacrifices the capacity of audio codec. Besides, the dependency among features are exploited at the receiver side only. In contrast, SoundSpring models the contextual relationship of scalable audio representations, benefiting both the efficiency and resiliency of the audio transceiver.

\vspace{-.05in}
\subsection{Objective Evaluation Results}

\begin{figure*}
	\centering
	\subfigure[]{
		\includegraphics[width=0.99\columnwidth]{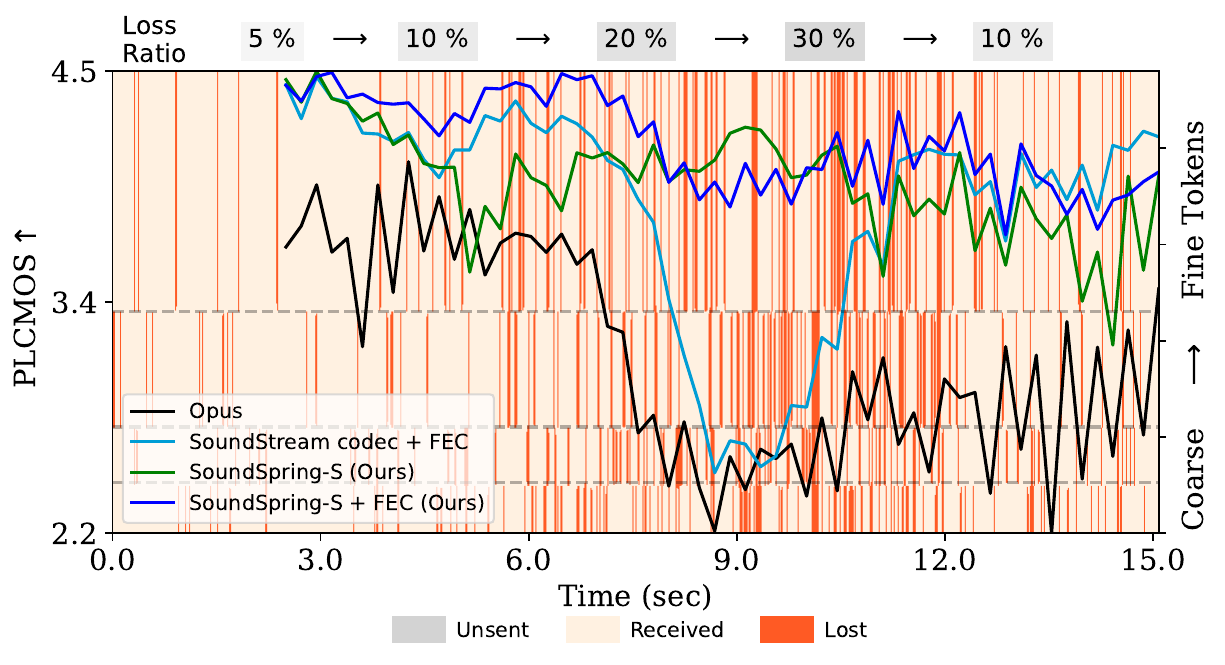}
	}
	\hspace{-.2in}
	\vspace{-.05in}
	\quad
	\subfigure[]{
		\includegraphics[width=0.99\columnwidth]{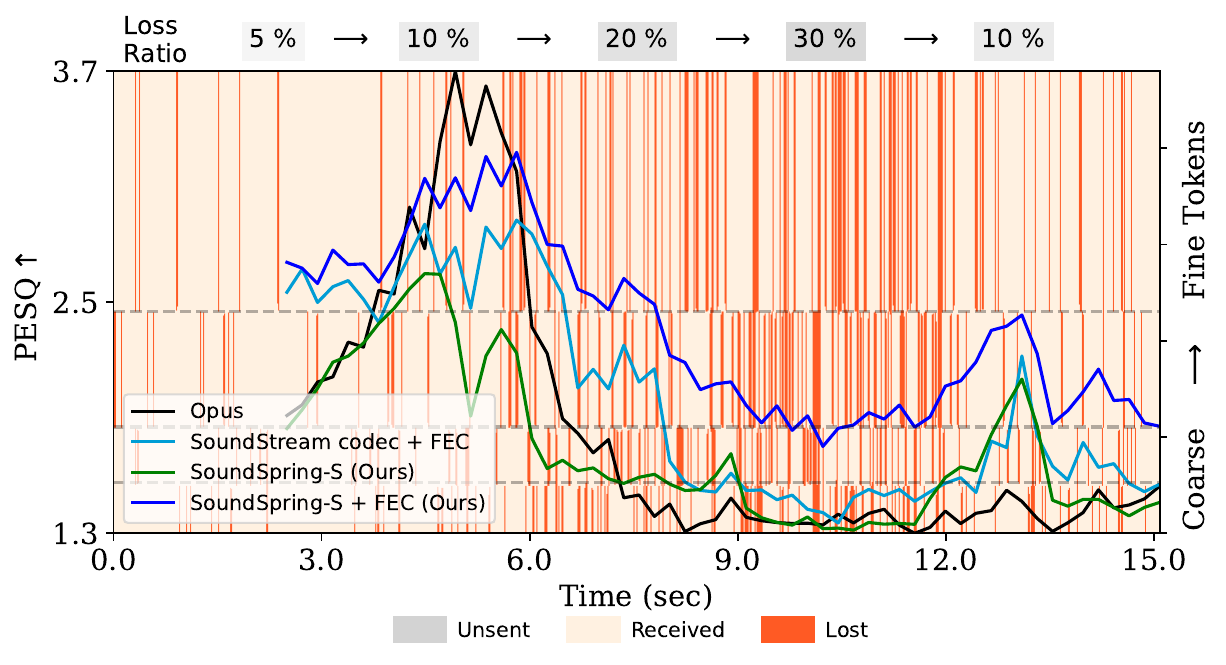}
	}
	
	\subfigure[]{
		\includegraphics[width=0.99\columnwidth]{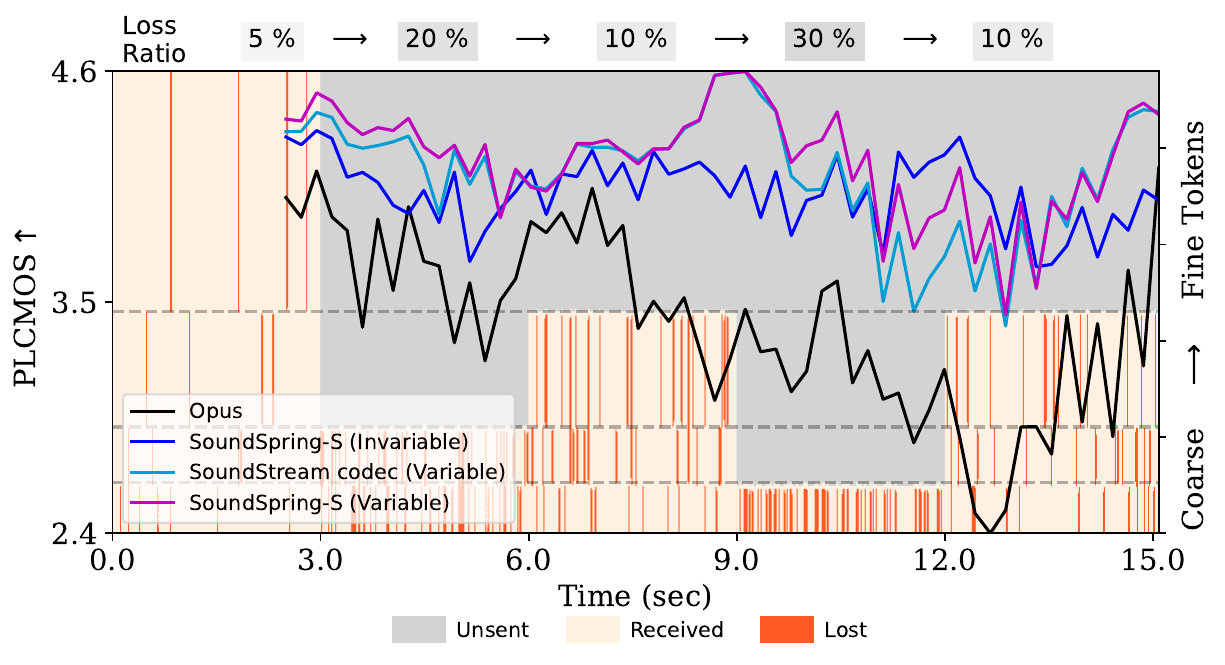}
	}
	\hspace{-.2in}
	\vspace{-.05in}
	\quad
	\subfigure[]{
		\includegraphics[width=0.99\columnwidth]{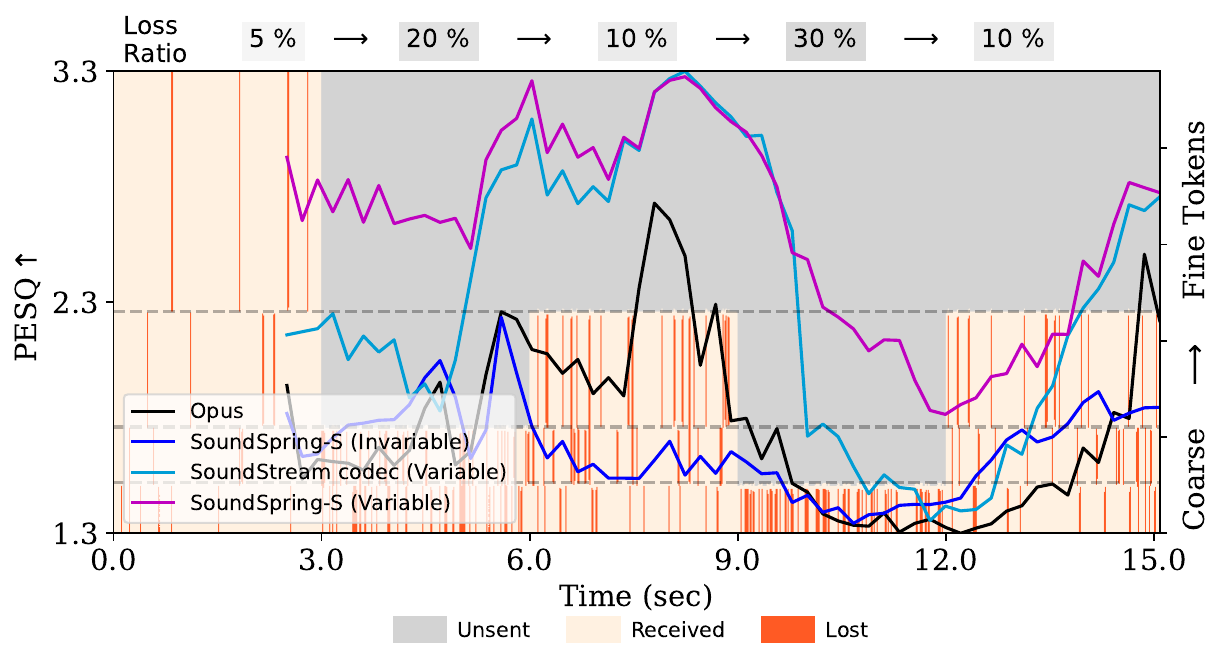}
	}
	\caption{Perceptual qualities of real-time speech transmission over variable-loss channels. Results in the top row (a)-(b) are tested with consistent $K=24$ layers of tokens are encoded and sent.
		Results in (c)-(d) are tested employing flexible coding strategy with variable $K \in \{3,6,12,24\}$. The total coding rate is close to ``SoundSpring-S (Invariable)'' with consistent $K=12$.
		FEC redundancy of coarse tokens is added in all schemes.
		The tokens are arranged as the background of the graph, with gray ones indicating unsent tokens and red marks indicating lost tokens.
		At each step, the score is reported by evaluating the latest 2.5-second audio segment.}
    \label{fig_variable_loss}
\end{figure*}

\subsubsection{Metrics}

We employ the widely-used Scale-Invariant Signal-to-Noise Ratio (SI-SNR)~\cite{sisnr}, and Signal-to-Distortion Ratio (SDR)~\cite{sdr} to measure the audio signal fidelity.

For speech signals specifically, there are also some objective perceptual quality metrics. We evaluate the difference of the mel-frequency cepstral coefficients (MFCC), which is the hand-crafted feature for automatic speech recognition. The MFCC difference is computed using the distance metric
\begin{equation}
 d_{\mathrm{MFCC}} = \frac{1}{F}\sum_{f=1}^F{\left\| m_f\left( \x \right) -m_f\left( \hat{\x} \right) \right\| _{2}^{2}},
\end{equation}
where $m_f$ is the MFCC function of $f$-th filterbank and $F$ is the number of mel scales (set to 4) with coefficient numbers of $[8, 16, 32, 64]$.
Besides, we use perceptual evaluation of speech quality (PESQ)~\cite{pesq} and ViSQOL~\cite{chinen2020visqol} as estimators of perceptual qualities of speech and audio, respectively.
PESQ, ranging from 1.0 to 4.5, is an ITU-T recommendation in P.862 to assess the speech quality of telephone networks and speech codecs when reference signals are available.
ViSQOL, ranging from 1 to 5, is also an objective and full-reference metric for perceived audio quality.

To evaluate the concealment capabilities, we leverage PLCMOS~\cite{plcmos}, a non-intrusive data-driven estimator of the mean opinion score for the evaluation of PLC algorithms.

\subsubsection{Performances on music dataset}

Performance on stereo music data over a 10\% packet loss channel is summarized in Table~\ref{tab_music}. We use the periodical slice segmentation as defined in \eqref{eq_period_slice} with $S=10$ slices for each 1-second GoS, i.e., $T_G=150$ frames, delivering tens of packets per second depending on the number of coding layers $K$.
Specifically, FEC redundancy of $N_C=2$ layers of coarse tokens introduces a 2.92 kbps overhead in average.
It can be observed that the dual-functional MLM demonstrates improvement in both compression and concealment performance in terms of distortion and perceptual metrics by leveraging the temporal and layer-wise context effectively.
Compared to traditional audio codecs (Opus and AAC), SoundSpring empowers the neural discrete audio representations with resilience, while maintaining or even improving coding efficiency compared to latent space quantization.
The reconstructed audio perceptually resembles the original one with lossless transmission, as indicated by ViSQOL metrics.
ViSQOL score is reported after remixing an 8-second 48kHz stereo audio to a mono one.

\subsubsection{Performances on speech dataset}

Speech quality versus data missing rate (DMR) performance is displayed in Fig.~\ref{fig_dmr}. For SoundSpring, FEC redundancy of coarse tokens is added to the subsequent frame. The overall coding rate combining redundancy is 5.0 kbps with a fixed $K=12$.
LBRR is adopted in ``Opus 18kbps'' as the FEC solution for Opus, while ``Opus 9kbps'' disables LBRR because the bitrate does not satisfy the requirement. Both Opus schemes employ frame interpolation for post-processing concealment.

It can be observed from Fig.~\ref{fig_dmr} that, even with a lower coding rate, SoundSpring outperforms conventional audio codecs with loss concealment across different loss ratios, in terms of distortion and perceptual quality metrics.
SoundSpring-S with strict causal audio coding and MLM modeling, exhibits a little performance degradation compared to SoundSpring with bi-directional context, yet it still outperforms the baselines without loss concealment. For instance, the PESQ performance of SoundSpring-S with 5 kbps is comparable to that of ``Opus 18kbps'' and superior to that without concealment when loss ratio is 30\%.
RVQ combined with post-processing FD-PLC, which is joint optimized with the audio decoder and evaluated in streaming scenarios, somehow has a negative impact on the capacity of audio codec. An evident performance degradation can be observed when the loss ratio is low, compared to that without concealment.

\subsubsection{Results under varying channel states}

Aside from evaluating the audio in a stationery channel environment, we test the real-time performance over channel with varying packet losses.
Fig.~\ref{fig_variable_loss} displays the real-time audio quality for SoundSpring-S with respect to PLCMOS and PESQ metrics under varying channel states.
All methods are tested using the same packet loss trace, which is randomly generated.
The trace of loss ratio versus time is marked above each subfigure.

Fig.~\ref{fig_variable_loss}(a) and Fig.~\ref{fig_variable_loss}(b) show the performance of two metrics with consistent $K=24$ layers of tokens encoded and transmitted (with yellow background in the graph).
It can be observed in Fig.~\ref{fig_variable_loss}(a), with the help of audio token concealment, SoundSpring-S with or without FEC can achieve stable PLCMOS performance, while the others' present U-shape curves encountering a period of long-term packet loss. Similarly, in Fig.~\ref{fig_variable_loss}(b), the PESQ performance of SoundSpring surpasses that of methods without concealment.

Besides, we verify the effectiveness of SoundSpring-S with variable-rate coding (by adjusting the number of token layers $K$) in Fig.~\ref{fig_variable_loss}(c) and Fig.~\ref{fig_variable_loss}(d), where we specify four levels of coding rate with $K=3,6,12,24$ each. Compared to invariable encoding with $K=12$, which has a similar bitrate in total, ``SoundSpring-S (Variable)'' has the potential of reducing the rate of the non-FEC portion and increasing the success rate of the restoration of coarse tokens. Therefore, ``SoundSpring-S (Variable)'' achieves a better balance in loss resilience between the loss concealment and FEC redundancy.
In terms of PESQ metric (Fig.~\ref{fig_variable_loss} (d)), the minimum real-time score of ``SoundSpring-S (Variable)'' is reported as around 1.9 while scores of others drop to around 1.4.

\begin{figure*}
	\setlength{\abovecaptionskip}{0.cm}
	\setlength{\belowcaptionskip}{-0.cm}
	\centering
	\includegraphics[scale=0.49]{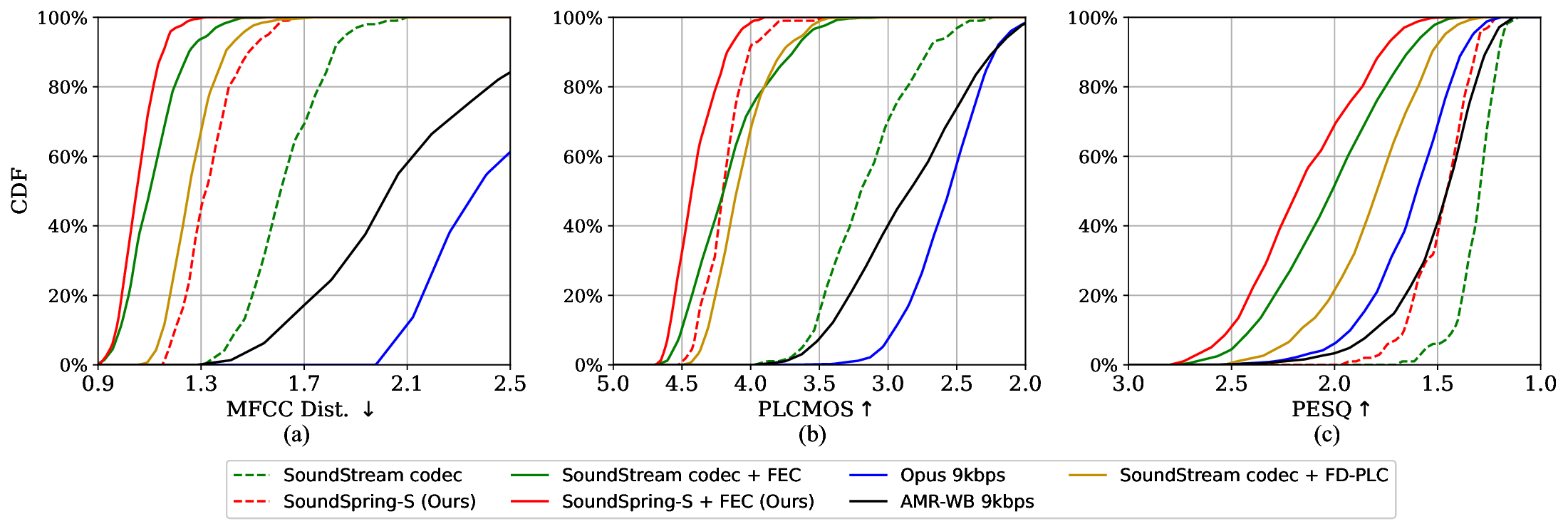}
	\caption{Cumulative distribution function (CDF) of audio quality metrics over a WLAN packet loss channel. The same $K$ layers of audio tokens are sent for all RVQ schemes, with coding rate of less than 6 kbps. The upward arrow $\uparrow$ indicates that higher values of the metric are favorable, and vice versa.
The x-axis is rearranged such that the CDF curve closer to the upper left of the graph indicates superior performance.}
	\label{fig_cdf}
\end{figure*}

\begin{figure*}
		\setlength{\abovecaptionskip}{0.cm}
	\setlength{\belowcaptionskip}{-0.cm}
  \centering
  \includegraphics[scale=0.5]{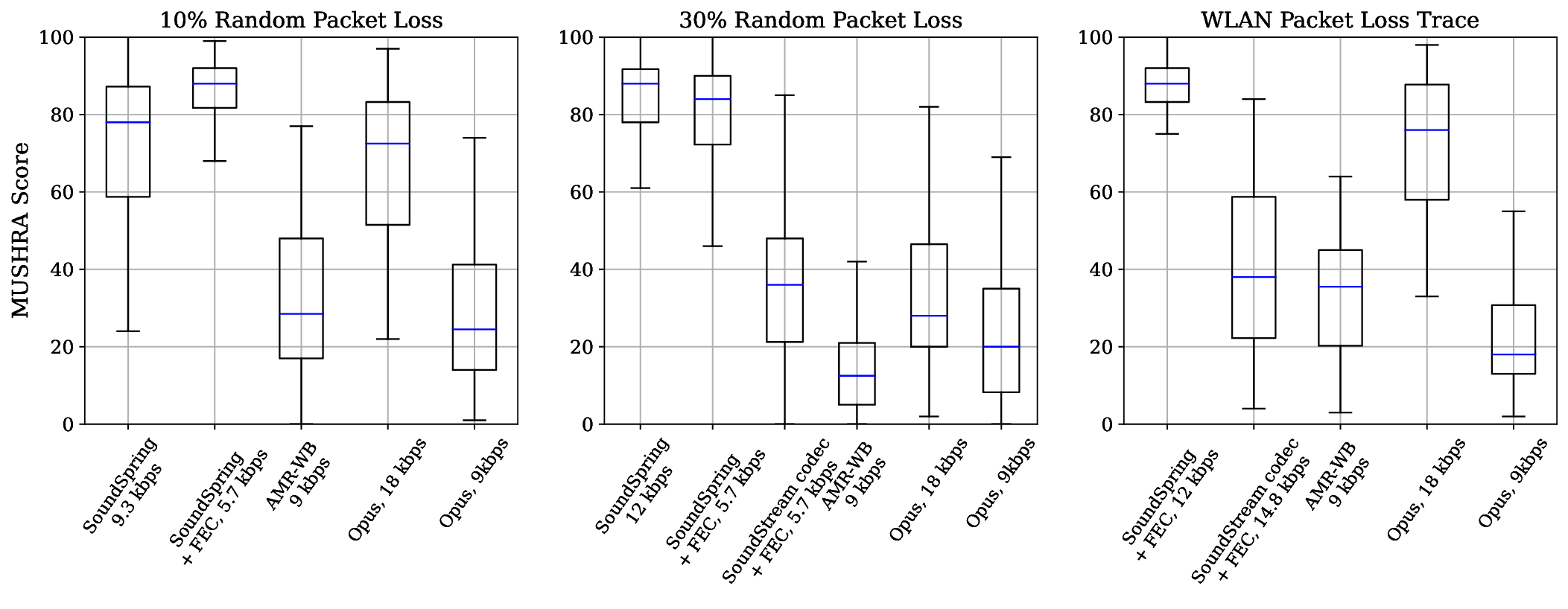}
  \vspace{-1em}
  \caption{MUSHRA scores simulated over channels with (a) 10\% random packet loss, (b) 30\% random packet loss, (c) WLAN packet loss trace. We provide demo examples of reconstructed audio for comparison at \href{https://semcomm.github.io/ResilientAudio}{https://semcomm.github.io/ResilientAudio}.}
  \label{fig_mushra}
\end{figure*}

Aside from independent and identically distributed (i.i.d.) random loss using Bernoulli model, it is expected that the proposed SoundSpring can handle the burst losses.
We conduct a large amount of simulations over the same random-selected audio clip by generating multiple packet loss traces using three-state Markov model~\cite{milner2004}, to simulate a WLAN channel with burst packet loss.
The resulting cumulative distribution function (CDF) of the quality metrics is shown in Fig.~\ref{fig_cdf}.
It can be observed that SoundSpring-S outperforms traditional audio codecs in terms of these perceptual metrics by a large margin.
The spectral feature of the audio concealed by SoundSpring is more close to the raw audio~(Fig.~\ref{fig_cdf}(a)).
The MLM concealment of SoundSpring effectively enhances the audio quality over lossy transmission, whether FEC is available or not.
In case of long burst losses, the FEC redundancy is more likely to lose effect.
On the contrary, the MLM concealment in SoundSpring is less affected by continuous packet loss. For example, given the same ratio of FEC redundancy, the percentage of audio with PLCMOS $>3.5$ increases from 20\% to nearly 100\% (Fig.~\ref{fig_cdf}(b)) by means of MLM audio concealment, and the percentage with PESQ $>2.0$ increases from 50\% to 70\% (Fig.~\ref{fig_cdf}(c)).
Despite less coding rate and the lack of FEC of coarse tokens (red dashed line), SoundSpring-S still achieves comparable PESQ performance with traditional audio codecs (Opus and AMR-WB at 9 kbps), which also have concealment capabilities through post-processing.

For SoundSpring, the contextual modeling at the sender compromises a certain degree of resilience, but yields compression gain.
When the same layers of audio tokens are transmitted, FD-PLC~\cite{msra_plc}, which integrates a post-processing concealing module before recovering audio, conceals better than SoundSpring without FEC protection.
However, SoundSpring combined with FEC redundancy with a total of 5.0 kbps is superior to FD-PLC at 6.0 kbps.
Given a neural audio codec, the MLM of SoundSpring facilitates a plug-and-play control of efficiency-resiliency balance.
Instead, the FD-PLC actually reshapes the audio codec, whose concealing performance is found to rely on the joint optimization with the neural audio codec with random dropout according to our experiments.

\subsection{Subjective Evaluation Results}

Aside from the established objective perceptual quality evaluation metrics, we also implement Multi-Stimulus Test with Hidden Reference and Anchor (MUSHRA) subjective test~\cite{MUSHRA} to verify our proposed system with human raters.
MUSHRA protocol allows users to compare and rate the perceived quality of different variants of the same audio clip, which are encoded and transmitted in different ways.
We randomly select 10 samples from the test set and each sample is rated by more than 20 annotators in three different channel settings, 10\% random packet loss, 30\% random packet loss, and the WLAN channel. To ensure fair comparisons, the packet loss trace is randomly sampled and keep the same for all variants.

Rating results in Fig.~\ref{fig_mushra} demonstrate that with token concealment, SoundSpring substantially improves the perceptual quality, achieving comparable or even better Mean Opinion Scores (MOS) compared to traditional codecs with FEC and higher coding rate. Meanwhile, when prediction errors occur during the concealment of coarse tokens, FEC for coarse tokens is an effective approach for auditory artifact suppression, which is more significant than a higher-quality restoration of fine tokens.
In case of long burst loss (Fig.~\ref{fig_mushra}(c)), the redundancy coding loses effect and incurs an overhead cost compared to MLM concealment. By incorporating the capability of concealing any loss pattern, MLM concealment together with FEC establishes a dual assurance mechanism for end users' QoE.

\begin{table}[t]
    \renewcommand{\arraystretch}{1.2}
    \centering
    \normalsize
    \caption{Real-time factor (RTF) w.r.t configuration of inference of SoundSpring-S.}
    \begin{adjustbox}{width=\columnwidth}
    \begin{tabular}{!{\vrule width1pt}m{.15\columnwidth}|m{.13\columnwidth}|m{.18\columnwidth}|m{.25\columnwidth}|m{.1\columnwidth}!{\vrule width1pt}}

    \Xhline{1pt}
    \centering Packet Loss (\%) & \centering Stride (ms) & \centering Contextual frames $T_C$ & \centering RTF (entropy modeling only) & \centering RTF (dual) \tabularnewline
    \Xhline{1pt}

    \centering 5 & \centering 60 & \centering 15 & \centering 0.36 & \centering 0.45 \tabularnewline
    \centering 10 & \centering 60 & \centering 15 & \centering 0.36 & \centering 0.51  \tabularnewline
    \centering 10 & \centering 100 & \centering 15 & \centering 0.26 & \centering 0.38 \tabularnewline
    \centering 10 & \centering 100 & \centering 30 & \centering 0.30 & \centering 0.42 \tabularnewline
    \centering WLAN  & \centering 100 & \centering 15 & \centering 0.26 & \centering 0.33 \tabularnewline
    \centering WLAN  & \centering 100 & \centering 30 & \centering 0.30 & \centering 0.39 \tabularnewline
    \centering WLAN  & \centering 200 & \centering 15 & \centering 0.14 & \centering 0.21 \tabularnewline

    \Xhline{1pt}

    \end{tabular}
    \end{adjustbox}
    \label{tab_latency}
\end{table}

\subsection{Latency Analysis}

End-to-end real-time factor (RTF) results are reported in Table~\ref{tab_latency}. RTF is defined as the ratio between the time needed for audio encoding/decoding and MLM inference and the audio duration. RTF $< 1$ indicates that the system can operate in real time.
Specifically, we evaluate the RTF of SoundSpring-S with MLM for entropy modeling only, and one with dual-functional MLM (the ``RTF (dual)'' column) in the table. The runtime of MLM is influenced by the concealing frequency, which is related to the packet loss and the number of contextual frames $T_C$.
Meanwhile, a longer stride $T_S$ between each inference increases the latency, and we restrict it within hundreds of milliseconds.

Results indicate that, a longer context length or a worse channel environment increases the loss rate of fine tokens, thus requiring more concealing operations at the receiver.
A longer stride of MLM inference yields a lower RTF at the sacrifice of a longer maximum latency.
We profiled all models on Intel Xeon Gold 6226R CPU with 16 kHz input for SoundSpring-S. 

\begin{table}[h]
\renewcommand{\arraystretch}{1.3}
\begin{center}
    \normalsize
    \caption{Real-time factor (RTF) and audio quality comparison with different network configurations of SoundSpring-S.}
    \begin{adjustbox}{width=1.0\columnwidth}
    \begin{tabular}{!{\vrule width1pt}m{.29\columnwidth}|m{.17\columnwidth}|m{.14\columnwidth}|m{.19\columnwidth}|m{.14\columnwidth}!{\vrule width1pt}}
    \Xhline{1pt}
    \centering Model Config & \centering Bitrate (kbps) & \centering PESQ $\uparrow$ & \centering \#Params of MLM (million)  & \centering RTF (dual) $\downarrow$ \tabularnewline
    \Xhline{1pt}
    \centering L12 D512 & \centering 5.0 & \centering 2.87 & \centering 81 & \centering 0.38 \tabularnewline
    \centering L3 D512 & \centering 5.1 & \centering 2.82 & \centering 59 & \centering 0.17 \tabularnewline
    \centering L6 D256 & \centering 5.2 & \centering 2.74 & \centering 28 & \centering 0.15  \tabularnewline
    \hline
    \centering SoundStream & \centering 6.0 & \centering 2.70 & \centering - & \centering 0.03  \tabularnewline
    \Xhline{1pt}
    \end{tabular}
    \label{tab_lite_model}
    \end{adjustbox}
\end{center}
\footnotesize{$^1$ Evaluation setup: 10\% random packet loss with FEC, $T_C=15$ contextual frames and 100ms stride.} \\
\footnotesize{$^2$ ``L12 D512'' denotes 12 Transformer layers with 512 dimensions.}
\vspace{-0.1in}
\end{table}

Furthermore, we investigate the impact of MLM configuration of SoundSpring-S on the audio transmission performance as well as the real-time property in Table~\ref{tab_lite_model}. By employing models with fewer layers or dimensions, the capability of MLM contextual modelling is weakened, leading to both coding rate increment and worse audio concealing performance.
We can also note a significant complexity increasing compared to SoundStream audio codec, due to the employment of high-dimensional LM. The optimization towards faster inference is required before the promising application of SoundSpring, by integrating high-speed LM inference techniques~\cite{liu2024lightweight}, e.g., caching, model pruning and quantization.

SoundSpring-S presents an initial trial to integrate real-time entropy modeling and token concealment by masked language modeling.
We acknowledge that traditional audio post-processing techniques introduce a lower latency and complexity, compared to learning-based solutions.
An appropriate network configuration should be chosen, depending on the computing capabilities of devices.
Possible joint efforts with these traditional methods as well as inference-efficient model development will catalyze its application in learning-based audio communication systems.

\section{Conclusion}\label{section_conclusion}

In this paper, we have introduced a novel audio transceiver, \emph{SoundSpring}, which integrates a powerful neural audio codec together with a dual-functional MLM.
Extensive results show that, the proposed SoundSpring has achieved a commendable balance between efficiency and resiliency against packet loss with no need of retransmission, especially in long burst packet loss scenarios, thus catalyzing many future RTC applications.
Our audio transceiver can conveniently plug-and-play with existing communication systems, thus promoting its practical deployment.
Hence, this work takes an important step towards future LLM-empowered intelligent and resilient transceivers.

\ifCLASSOPTIONcaptionsoff
  \newpage
\fi

\bibliographystyle{IEEEtran}
\bibliography{Ref}

\vfill
\end{document}